# Quantum optimal control of steady orbits

Shebha Anandhi Jegadeesan[1,#], Yujie Zhao[2,#], Robert Hunter[2], Hassane El Mkami[2], Maximilian Keitel[3], Callum Musselwhite[3], Graham Smith[2,*], Guinevere Mathies[1,*], Ilya Kuprov[4,*]

[1]*Department of Chemistry, University of Konstanz, Konstanz, Germany.*

[2]*School of Physics and Astronomy, University of St Andrews, St Andrews, United Kingdom.*

[3]*School of Chemistry and Chemical Engineering, University of Southampton, Southampton, United Kingdom.*

[4]*Department of Chemical and Biological Physics, Weizmann Institute of Science, Rehovot, Israel.*

## Abstract

Periodically driven dissipative systems can settle into steady orbits – fixed infinite loops on their dynamical manifolds. In quantum mechanics, steady orbits occur in cooling engines, lasers and masers, precision metrology devices, and magnetic resonance. Steady orbits and stroboscopic steady states are a promising target for quantum optimal control, but the numerical complexity is prohibitive: infinite loops defeat state-of-the-art algorithms that use explicit numerical propagation in the time domain. We report a solution to this problem and introduce a numerically efficient quantum optimal control algorithm for steady orbits. We use this algorithm to design microwave pulse sequences that dynamically polarise nuclear spins and enhance their magnetic resonance signals, while taking into account instrumental distortions. Improved performance relative to leading current methods is demonstrated experimentally.

*Email: *gms@st-andrews.ac.uk*
*guinevere.mathies@uni-konstanz.de*
*ilya.kuprov@weizmann.ac.il*

[#]These authors contributed equally.

# 1. Introduction

Periodically driven systems rarely settle into a static point. Instead, in the presence of damping, they may approach a periodic attractor – a "stroboscopic" steady state and its associated semi-group orbit[1,2]. Such dynamics (drive + loss → limit cycle) are ubiquitous[3,4]: lasers and masers work using driven and damped light[5,6], optomechanical self-oscillators yield stable frequency references[7]; dissipative Kerr solitons enable integrated microcombs[8]; coherent population trapping and electromagnetically induced transparency enable precise clocks and magnetometers[9–11]; dynamic nuclear polarisation improves magnetic resonance sensitivity[12,13].

For asymptotically attracting steady orbits to exist, energy dissipation must be present. In that case, Schrödinger's equation is no longer sufficient; it must be upgraded to the thermodynamically consistent density operator formalism[14–16]. Mathematically, we are then looking for a fixed point density operator $\boldsymbol{\rho}_\infty$ satisfying $\mathbf{P}\boldsymbol{\rho}_\infty = \boldsymbol{\rho}_\infty$ for the time evolution superoperator $\mathbf{P}$ created by whatever the instrument repeatedly does, the technical term is "control sequence". Under the usual primitivity and ergodicity assumptions, such fixed points exist and are unique; their steady orbits are what physical systems settle into after many cycles of periodic driving[16,17].

A useful goal is to control steady orbits with the same care as quantum gates[18–20] and destination states[21–23]: the challenge then is not merely to prepare a given state or design a particular gate, but to engineer an entire limit cycle. This means so choosing a control sequence that its repeated application drives any initial state towards a periodic trajectory passing, with maximum robustness and fidelity, through user-specified waypoints (Figure 1). The special case of a single waypoint, usually the control loop endpoint, optimises a stroboscopic steady state.

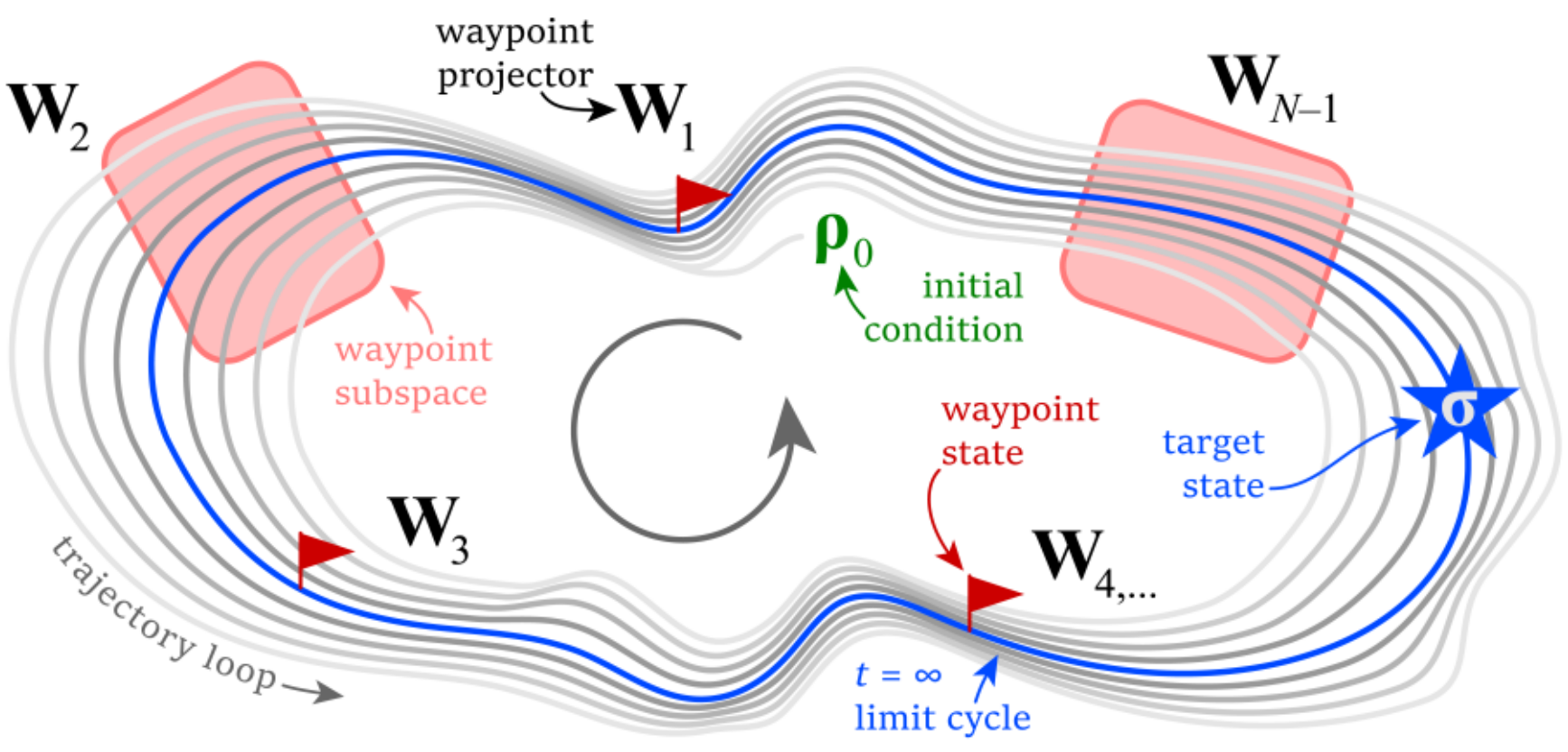

***Figure 1. Schematic illustration of steady orbit optimal control.*** *The system starts at the initial state $\boldsymbol{\rho}_0$ (green) and evolves (grey lines) under a periodic external control sequence. The original gradient ascent pulse engineering (GRAPE) method[24,25] designs a single-shot control sequence that takes the system to a specified destination state $\boldsymbol{\sigma}$ (blue star). The formalism proposed in this article designs instead a periodic control sequence that makes the system approach the blue attractor loop in the state space, constrained by waypoint projectors $\mathbf{W}_n$ enforcing a state (red flags) or a subspace (red areas).*

The prior art has so far targeted a different objective: in magnetic resonance and quantum technologies there are plenty of methods that yield effective Hamiltonians, unitary gates, and precise destination states after single or repeated control sequences, sometimes with robustness to noise and instrumental artefacts. None of the work reviewed below targets asymptotic limit cycles, but a lot has been done in related areas that merits an overview.

Much of the early work was done by the magnetic resonance community. Over twenty years ago, Nielsen group designed low-power dipolar recoupling sequences using optimal control with power constraints, but their aim was to make a specific effective Hamiltonian, not a limit cycle[26]. Their subsequent work on double-quantum dipolar recoupling also had an average Hamiltonian type unitary dynamics objective[27]. Optimal tracking from the Glaser group targeted unitary decoupling performance, also a property of the average Hamiltonian[28]. That was all done with gradient ascent pulse engineering (GRAPE) type algorithms[24,25,29]. In the variational corner[30,31], monotonically convergent NMR and DNP pulse design methods only considered destination state fidelity[32]. The latest proposals for optimal control strategies in dynamic nuclear polarisation[33,34] do not view relaxation as an optimisation objective and focus instead on the effective Hamiltonian design; a detailed review of the relevant literature was published by the Cory group[35].

Elsewhere in physics, robust Hamiltonian engineering for many-body systems with periodic control sequences is also established for unitary dynamics[36,37], objectives being state space transformation accuracy and robustness, with decoherence treated as a nuisance to be avoided rather than a feature to be used[37,38]. The well-travelled isomorphism between Lie algebras of all two-level systems has allowed magnetic resonance control theory to be ported to atom interferometers[39]. Dissipative dynamics come useful for state preparation after one or more control events ("stabiliser pumping"[40] in very recent literature), for example for resonator reset in circuit QED[41], but the target is again a state rather than orbit.

The common methodological thread in the prior art is Hamiltonian engineering: existing literature designs effective Hamiltonians and / or state transfer mechanisms, sometimes with robustness to classical or quantum mechanical imperfections. It does not formulate or solve the following problem: *find optimal controls such that the evolution has a specified attractive limit cycle*. In particular, there is no GRAPE type method for sequences that, when applied repeatedly, drive a dissipative quantum system towards a steady orbit that passes through user-specified waypoints and populates, to the maximum extent, a user-specified stroboscopic steady state.

In magnetic resonance, steady orbits exist in the Frydman pump[42] which transports nuclear magnetisation from solvent-exchangeable protons to the $^{13}C$ nuclei of the protein backbone. Another example is pulsed dynamic nuclear polarisation (DNP)[43], where a microwave control sequence is repeatedly applied to a sample doped with a paramagnetic species in order to transfer electron magnetisation to nuclei. Pulsed DNP is particularly effective in the solid state where dipole couplings mediate the magnetisation transfer; its combination with high-resolution magic-angle spinning (MAS) NMR at magnetic fields of up to 30 T would expand the characterisation toolkit for complex materials and biosolids[44–46]. Given inevitable limitations on the average and peak microwave power, it is desirable to optimise DNP experiments for efficiency. This is a stroboscopic steady state optimisation problem: electron spin relaxation is what drives nuclear hyperpolarisation – dissipative dynamics and the control sequence loop must be included in the modelling[47].

So far, the design of DNP pulse sequences has relied on the analysis of their single-shot application with no consideration of steady orbits. Experimental implementation has focused on low magnetic fields (up to 1.2 T) for which commercial electron spin resonance (ESR) instruments are available. Narrow-line organic radicals in glassy frozen solutions, with zero-quantum (ZQ) and double-quantum (DQ) transitions well separated from the single-quantum (SQ) transition, are convenient and practically useful test systems[48]. Using perturbation theory, one readily finds, for a microwave pulse of constant amplitude and phase[49], that electron magnetisation is transferred to nuclei when the effective nutation frequency $\sqrt{\omega_{1S}^2+\Omega^2}$ (where $\omega_{1S}$ is the electron Rabi frequency and $\Omega=\omega_{0S}-\omega_{\mu w}$ is the microwave resonance offset) matches the nuclear Larmor frequency. Nuclear spin orientation via electron spin locking (NOVEL)[50] and solid effect (SE)[51,52] DNP experiments use this matching condition. Integrated, stretched, and adiabatic solid effect (ISE, SSE, ASE)[53–57] experiments rely on the adiabatic passage[58] through the same condition. Several DNP pulse sequences with modulating amplitudes or phases were inspired by decoupling sequences from MAS NMR[59–62]. The matching condition of these sequences is easier to meet at higher magnetic fields, because it contains the modulation frequency. Of these, X-inverse-X (XiX)[60] is the most efficient when microwave peak power is limited. DNP sequences inspired by recoupling [63,64] and those designed with optimal control have also been tried[65,66], but they typically require impractically high peak microwave powers at high field.

In this communication, we report a mathematical formulation, a software implementation (into the *Spinach* library[67]), a hardware implementation (on the HiPER instrument[68]), and an experimental demonstration of steady orbit optimal control. The proposed algorithm has the same numerical complexity scaling as the classical GRAPE method[24]. We use it to design DNP pulse sequences for realistic conditions at 3.35 T (143 MHz proton and 94 GHz electron Larmor frequency) with restricted microwave power. Control field inhomogeneity and the filter function of the microwave electronics are incorporated into the design procedure. Compared to the targeted adiabatic passage technique ASE, a steady orbit GRAPE pulse sequence yields a small but significant improvement in NMR sensitivity.

## 2. Steady orbit GRAPE

We begin with the standard Liouville space (*aka* adjoint representation) formulation[25] of gradient ascent pulse engineering[24] in which the equation of motion for the density matrix is a dissipative generalisation of the Liouville – von Neumann equation:

$$\begin{gathered}\frac{\partial}{\partial t}\boldsymbol{\rho}(t)=-i\mathbf{L}(t)\boldsymbol{\rho}(t),\quad \mathbf{L}=\mathbf{H}+i\mathbf{R}+\ldots\\ \boldsymbol{\rho}(t+dt)=\exp\left[-i\mathbf{L}(t)dt\right]\boldsymbol{\rho}(t)\end{gathered}\tag{1}$$

where $\boldsymbol{\rho}$ is the density matrix, $\mathbf{H}$ is the Hamiltonian commutation superoperator and $\mathbf{R}$ is a relaxation superoperator driving the system towards the thermal equilibrium state. The evolution generator $\mathbf{L}$ can also have other terms – for example, kinetics, diffusion, hydrodynamics, and magic angle spinning[69].

Dissipative GRAPE framework[25,29] partitions $\mathbf{L}$ into the uncontrollable drift part $\mathbf{D}$ and a linear combination of control superoperators $\mathbf{C}_k$ whose coefficients $c_n^{(k)}$ the instrument can vary at each finite time slice $\Delta t_n$ in the discrete-time solution of the equation of motion:

$$\begin{gathered}\mathbf{L}_n = \mathbf{D}_n + \sum_k c_n^{(k)} \mathbf{C}_k \\ \mathbf{P}_n = \exp\left[-i\mathbf{L}_n \Delta t_n\right], \quad \boldsymbol{\rho}_n = \mathbf{P}_n \boldsymbol{\rho}_{n-1}\end{gathered} \tag{2}$$

where $\mathbf{P}_n$ is the propagator of the *n*-th discrete time slice. The fidelity in the GRAPE framework is a function, sometimes an elaborate one[70–74], of the overlap between the system state at the end of the one-off control sequence and some desired target state:

$$\Omega_1 = \langle \boldsymbol{\sigma} | \mathbf{P}_N \mathbf{P}_{N-1} \cdots \mathbf{P}_2 \mathbf{P}_1 | \boldsymbol{\rho}_0 \rangle \tag{3}$$

where $\boldsymbol{\rho}_0$ is the initial density matrix, $\boldsymbol{\sigma}$ is the target density matrix, the Frobenius inner product is $\langle \mathbf{A} | \mathbf{B} \rangle = \mathrm{Tr}\left(\mathbf{A}^\dagger \mathbf{B}\right)$, $N$ is the number of discrete time steps, and the index in $\Omega_1$ refers to the fact that the control sequence is here applied only once.

To encourage trajectory passage through specific subspaces or states at specific times, we insert waypoint projectors into the evolution rule and its subsequent discretisation[29]:

$$\begin{gathered}\boldsymbol{\rho}(t+dt) = \mathbf{W}(t+dt)\exp\left[-i\mathbf{L}(t)\,dt\right]\boldsymbol{\rho}(t) \\ \Downarrow \\ \boldsymbol{\rho}_n = \mathbf{W}_n \mathbf{P}_n \boldsymbol{\rho}_{n-1}, \quad \mathbf{W}_n = \mathbf{W}(t_n)\end{gathered} \tag{4}$$

where $\mathbf{W}(t)$ is the projector into the subspace or the state where we require the system to be at time $t$. When the system does not pass through that subspace or state, the projectors bite and the fidelity is reduced. Conveniently, $\mathbf{W}_n$ are constant Hermitian matrices (supplied to *Spinach* in practice as linear idempotent function handles) – they do not interfere with the subsequent GRAPE mathematics. We would therefore not mention them each time but simply imply their possible presence after each time step propagator. However, waypoint projectors are only a mathematical trick: when a physical trajectory or a steady state are computed, they must be omitted; all of this is implemented in *Spinach*[29].

GRAPE is popular because the gradient of $\Omega_1$ with respect to the control sequence:

$$\frac{\partial \Omega_1}{\partial c_n^{(k)}} = \langle \boldsymbol{\sigma} | \mathbf{P}_N \mathbf{P}_{N-1} \cdots \frac{\partial \mathbf{P}_n}{\partial c_n^{(k)}} \cdots \mathbf{P}_2 \mathbf{P}_1 | \boldsymbol{\rho}_0 \rangle \tag{5}$$

may be computed very efficiently[24,75], and even analytically for simple systems[76]. There is a vibrant industry of options, algorithms, implementations, and applications; instrumental distortions of control sequences can also be accounted for[77].

Having reviewed the settings, we now consider the stroboscopic steady state as the optimisation target. The control sequence is now applied to the initial condition repeatedly; the quantity of interest therefore acquires an infinite loop:

$$\Omega_\infty = \langle \boldsymbol{\sigma} | \boldsymbol{\rho}_\infty \rangle, \quad |\boldsymbol{\rho}_\infty\rangle = \lim_{m\to\infty} \left(\mathbf{P}_N \mathbf{P}_{N-1} \cdots \mathbf{P}_2 \mathbf{P}_1\right)^m |\boldsymbol{\rho}_0\rangle \tag{6}$$

where the result may not actually depend on $\boldsymbol{\rho}_0$ even though it is nominally present[47]. This is a difficult optimisation target: even if the limit is truncated at some finite $m$, propagators would not usually commute with their derivatives – the gradient of $\Omega_{m\gg 1}$ with respect to the control sequence would contain expensive product rule expansions. That is the route taken for unitary dynamics by Carvalho *et al.*[33] who have used an auxiliary matrix relation[78]:

$$f\left[\begin{pmatrix} \mathbf{A} & \partial\mathbf{A}/\partial\alpha \\ \mathbf{0} & \mathbf{A} \end{pmatrix}\right]=\begin{pmatrix} f(\mathbf{A}) & \partial f(\mathbf{A})/\partial\alpha \\ \mathbf{0} & f(\mathbf{A}) \end{pmatrix} \tag{7}$$

to differentiate propagator powers and logarithms; the latter were used to obtain effective Hamiltonians. The method worked well for the approximate unitary models of DNP experiments considered in that paper. However, under dissipative dynamics (a strict requirement for quantitative models of pulsed DNP[47]) this is a perilous and resource-intensive approach: dissipative propagator logarithm is numerically unstable and may not exist to begin with. Even when it does, it has cubic complexity with no efficient sparse, parallel, or GPU implementations because matrix factorisations are required for computing sequential square roots[79].

In view of the above, we propose a different route towards optimal control of steady orbits and stroboscopic steady states. It uses the recently designed Newton-Raphson type method[47] for computing them. Consider the definition of the stroboscopic steady state $\boldsymbol{\rho}_\infty$

$$\boldsymbol{\rho}_\infty=\left(\mathbf{P}_N\mathbf{P}_{N-1}\cdots\mathbf{P}_2\mathbf{P}_1\right)\boldsymbol{\rho}_\infty \tag{8}$$

used by the Newton-Raphson solver.[47] When this is used instead of Eq (6), $\Omega_\infty$ becomes much easier to differentiate because the infinite loop is gone:

$$\begin{gathered}\Omega_\infty=\langle\boldsymbol{\sigma}|\boldsymbol{\rho}_\infty\rangle,\qquad \frac{\partial\Omega_\infty}{\partial c_n^{(k)}}=\left\langle\boldsymbol{\sigma}\middle|\frac{\partial\boldsymbol{\rho}_\infty}{\partial c_n^{(k)}}\right\rangle \\ \frac{\partial\boldsymbol{\rho}_\infty}{\partial c_n^{(k)}}=\left[\frac{\partial}{\partial c_n^{(k)}}\left(\mathbf{P}_N\mathbf{P}_{N-1}\cdots\mathbf{P}_2\mathbf{P}_1\right)\right]\boldsymbol{\rho}_\infty+\left(\mathbf{P}_N\mathbf{P}_{N-1}\cdots\mathbf{P}_2\mathbf{P}_1\right)\frac{\partial\boldsymbol{\rho}_\infty}{\partial c_n^{(k)}}\end{gathered} \tag{9}$$

In the latter equation, the first term on the right hand side comes at no extra cost relative to standard GRAPE because it is already computed as a part of the GRAPE algorithm workflow[25,29]:

$$\left[\frac{\partial}{\partial c_n^{(k)}}\left(\mathbf{P}_N\mathbf{P}_{N-1}\cdots\mathbf{P}_2\mathbf{P}_1\right)\right]\boldsymbol{\rho}_\infty=\left(\mathbf{P}_N\mathbf{P}_{N-1}\cdots\frac{\partial\mathbf{P}_n}{\partial c_n^{(k)}}\cdots\mathbf{P}_2\mathbf{P}_1\right)\boldsymbol{\rho}_\infty \tag{10}$$

After a straightforward rearrangement, we obtain the steady state fidelity gradient:

$$\begin{gathered}\frac{\partial\Omega_\infty}{\partial c_n^{(k)}}=\langle\boldsymbol{\sigma}_\infty|\left(\mathbf{P}_N\mathbf{P}_{N-1}\cdots\frac{\partial\mathbf{P}_n}{\partial c_n^{(k)}}\cdots\mathbf{P}_2\mathbf{P}_1\right)|\boldsymbol{\rho}_\infty\rangle \\ \langle\boldsymbol{\sigma}_\infty|=\langle\boldsymbol{\sigma}|\left(\mathbf{1}-\mathbf{P}_N\mathbf{P}_{N-1}\cdots\mathbf{P}_2\mathbf{P}_1\right)^{-1}\end{gathered} \tag{11}$$

which is identical to the usual GRAPE fidelity gradient expression (and can therefore use all existing implementations) except for one local patch: the destination state needs to be modified by applying $\left(\mathbf{1}-\mathbf{P}\right)^{-1}$ where $\mathbf{P}=\mathbf{P}_N\mathbf{P}_{N-1}\cdots\mathbf{P}_2\mathbf{P}_1$ is the control sequence loop propagator without keyholes. This may be done using only matrix-vector products[80]. Dissipation is beneficial here because it moves propagator eigenvalues inside the unit circle and improves the stability of $\langle\boldsymbol{\sigma}|(\mathbf{1}-\mathbf{P})^{-1}$ operation. The trace eigenvalue of $\mathbf{P}$ (always strictly 1 to maintain the unit sum of probabilities) does not cause problems because the trace of $\boldsymbol{\sigma}$ is zero.

This procedure is distinct from Floquet and Floquet-Lindblad state engineering[39,40,81] because it targets (through waypoint projectors) a user-specified limit cycle: not just a generator or a stroboscopic eigenstate of the control sequence period map. The proposed algorithm is also computationally efficient (same numerical complexity scaling as GRAPE) because $\langle\boldsymbol{\sigma}|(\mathbf{1}-\mathbf{P})^{-1}$ may be computed using only sparse matrix-vector operations[82]. Dissipation here is not a bug, but a feature: it provides the contraction that makes the desired orbit an attractor.

# 3. Patching existing GRAPE implementations

Conveniently, the steady orbit version of GRAPE is a localised patch in any existing Liouville space GRAPE implementation[29]: the fidelity and the gradient are still to be returned, the user-specified initial condition is to be ignored because the steady orbit does not depend on it. In that way, all upstream functionality of the optimal control module of *Spinach*[67] remains the same, and only the lowest level `grape_liouv.m` function[29] is modified. Patch stages:

1. Compute the full propagator $\mathbf{P}=\mathbf{P}_N\mathbf{P}_{N-1}\cdots\mathbf{P}_2\mathbf{P}_1$ with waypoint projectors omitted and use the Newton-Raphson method[47] to obtain the stroboscopic steady state $|\boldsymbol{\rho}_\infty\rangle$.
2. Compute the modified destination state:

   $$\langle\boldsymbol{\sigma}_\infty| = \langle\boldsymbol{\sigma}|(\mathbf{1}-\mathbf{P})^{-1} \qquad \Rightarrow \qquad |\boldsymbol{\sigma}_\infty\rangle = (\mathbf{1}-\mathbf{P})^{-\dagger}|\boldsymbol{\sigma}\rangle \tag{12}$$

   The presence of dissipation makes this stage computable because all eigenvalues of $\mathbf{P}$ except the trace eigenvalue are inside the unit circle, and the trace of $\boldsymbol{\sigma}$ is zero.
3. Call regular GRAPE with $|\boldsymbol{\rho}_\infty\rangle$ as the initial condition, $|\boldsymbol{\sigma}\rangle$ as the destination state for the fidelity calculation, and $|\boldsymbol{\sigma}_\infty\rangle$ as the starting point for the backward trajectory; the propagators at this stage must include waypoint projectors:

   $$\begin{aligned} \Omega_\infty &= \langle\boldsymbol{\sigma}|\mathbf{P}_N\mathbf{P}_{N-1}\cdots\mathbf{P}_2\mathbf{P}_1|\boldsymbol{\rho}_\infty\rangle \\ \frac{\partial\Omega_\infty}{\partial c_n^{(k)}} &= \langle\boldsymbol{\sigma}_\infty|\mathbf{P}_N\mathbf{P}_{N-1}\cdots\frac{\partial\mathbf{P}_n}{\partial c_n^{(k)}}\cdots\mathbf{P}_2\mathbf{P}_1|\boldsymbol{\rho}_\infty\rangle \end{aligned} \tag{13}$$

This procedure is more expensive than ordinary Liouville space GRAPE by a constant factor (the propagator is formed explicitly once), but it retains the key advantage of that method – the gradient is cheap and the complexity scaling with the number of control sequence slices is linear.

A large number of computer science level optimisations usually applied to GRAPE (parallelisation, GPU processing, sparse matrix routines, tensor-structured methods, quasi-Newton and Newton-Raphson optimisation algorithms, *etc.*) remain applicable[29]. Annotated *Matlab* code is available in versions 3.0 and later of *Spinach* library (https://github.com/IlyaKuprov/Spinach)[67].

## 4. Pulsed DNP experiments

DNP experiments were performed at 3.35 T on a spectrometer known as HiPER. This instrument was originally developed at the University of St Andrews for pulse ESR with maximum concentration sensitivity and minimum dead time.[68] Capability of pulsed DNP has been demonstrated with HiPER[83] and with its twin[55] at the National High Magnetic Field Laboratory in Tallahassee. HiPER employs an extended-interaction klystron amplifier (EIKA, Communications and Power Industries Canada, Inc.) to amplify 94 GHz pulses into the kilowatt power range. Its non-resonant sample holder and rigorous suppression of standing waves yield a near-flat frequency response over a 1 GHz window. A recent upgrade of the microwave bridge has enabled arbitrary amplitude and phase modulation of high-power pulses at 94 GHz.[84]

A simplified scheme of HiPER is shown in Figure 2 (see Figure S1 for further details). An arbitrary waveform generator (AWG, Keysight Technologies, green in Figure 2) creates the desired waveform at 1.8 GHz with a time resolution of ~0.1 ns. The signal is then upconverted by mixing with 92.2 GHz (red) to 94 GHz, filtered, and sent through a rotary vane attenuator (RVA) before amplification by the EIKA (blue). Quasi-optical techniques are used to guide microwaves to and from the probe.[68] The sample holder is operated in the induction mode, where electron spins are excited with a linearly polarised beam and the induction signal is returned in an orthogonal mode; this removes the need for high-power receiver protection – a PIN switch suffices (black). Following downconversion to 1.8 GHz, amplification, and filtering (orange), quadrature detection is performed, with the AWG providing a coherent 1.8 GHz signal as a reference (lilac).

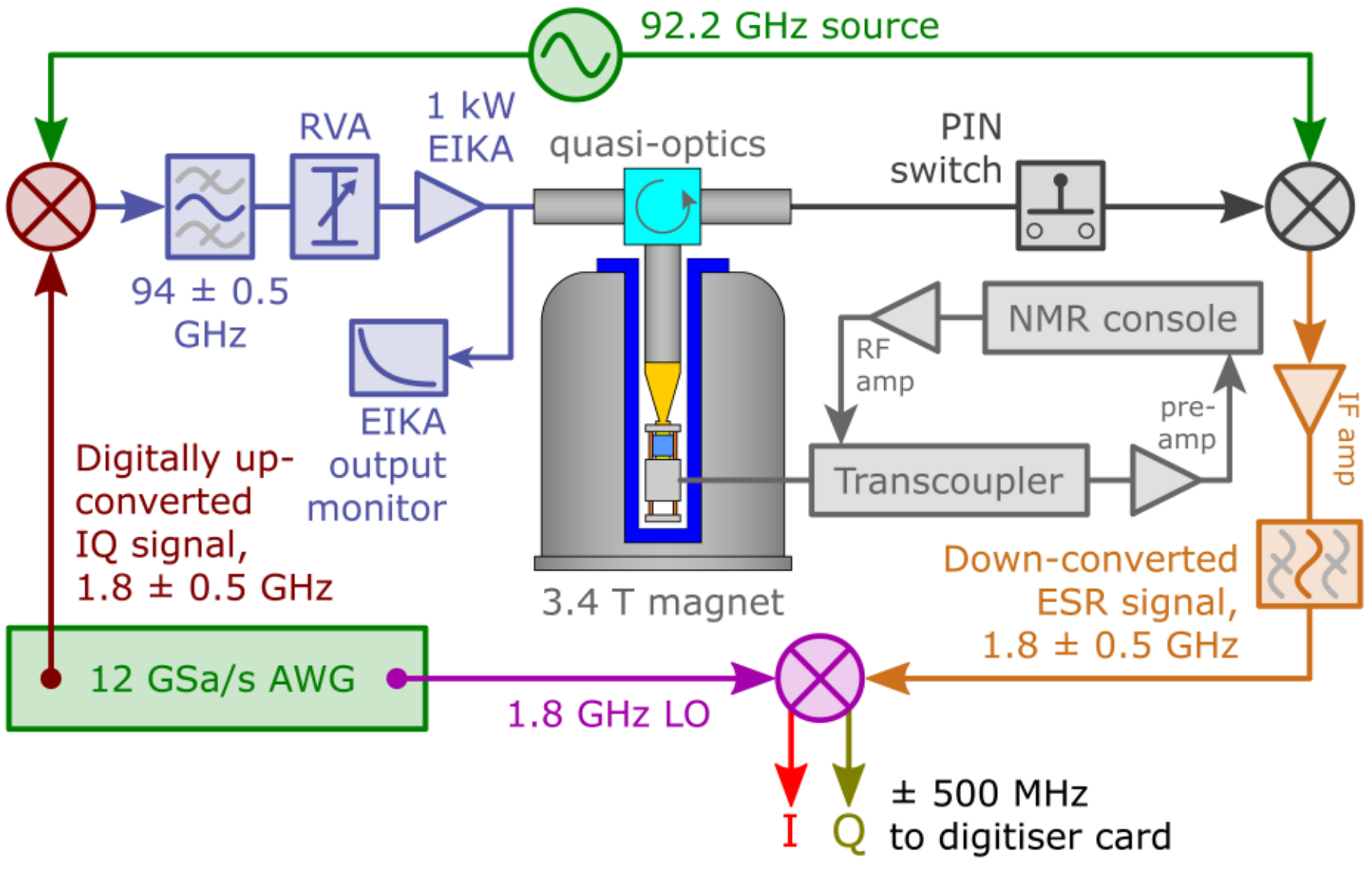

***Figure 2. Simplified schematic of the HiPER pulsed ESR/DNP spectrometer operating at 3.35 T.*** *Integration of an arbitrary waveform generator (AWG) into the heterodyning scheme enables convenient pulse shaping at 94 GHz and coherent quadrature detection of the ESR signal. For further details, see the text, the papers describing HiPER[68] and its upgrades[84], and Figure S1.*

In the standard HiPER pulsed ESR probe, a fluorinated ethylene propylene (FEP) sample tube (ID 2 mm, OD 3 mm) is placed inside a shorted circular metal waveguide that supports both linearly polarised orthogonal $TE_{11}$ modes. For pulsed DNP experiments[83], a portion of the metal waveguide is removed, but microwaves are still confined to the sample when it has a higher refractive index than the FEP tube.[84] This allows for efficient $^1$H NMR excitation and detection (90° pulse of 8 µs at 143 MHz) by a locally tuned and matched saddle coil across the active sample volume of 30 µL. A separate NMR circuit, comprising a console (Tecmag), power amplifier, transcoupler for signal routing, and receiver pre-amplifier (grey in Figure 2), handles the generation of radiofrequency pulses and nuclear induction signal acquisition.

The sample consisted of the narrow-line organic radical trityl OX063 (Polarize ApS) dissolved to a concentration of 5 mM in a glass-forming matrix of $d_8$-glycerol:$D_2O$:$H_2O$ in a 6:3:1 ratio by volume[48]. Before loading into the probe and cryostat, the solution was flash frozen in the FEP tube using liquid nitrogen. During experiments, the probe temperature was kept at 80 K using nitrogen as a cryogen. Prior to pulsed DNP experiments, the ESR spectrum of the trityl sample was measured by echo detection (Figure S2a). The full width at half-maximum of the ESR line was 52 MHz and the maximum echo intensity was observed at 94.004 GHz. Transverse electron magnetisation decay (characteristic times of 2.4 µs and 77 µs, Figure S2b) and longitudinal electron relaxation (characteristic times of 0.21 and 1.4 ms, Figure S2c) were both found to be biexponential.

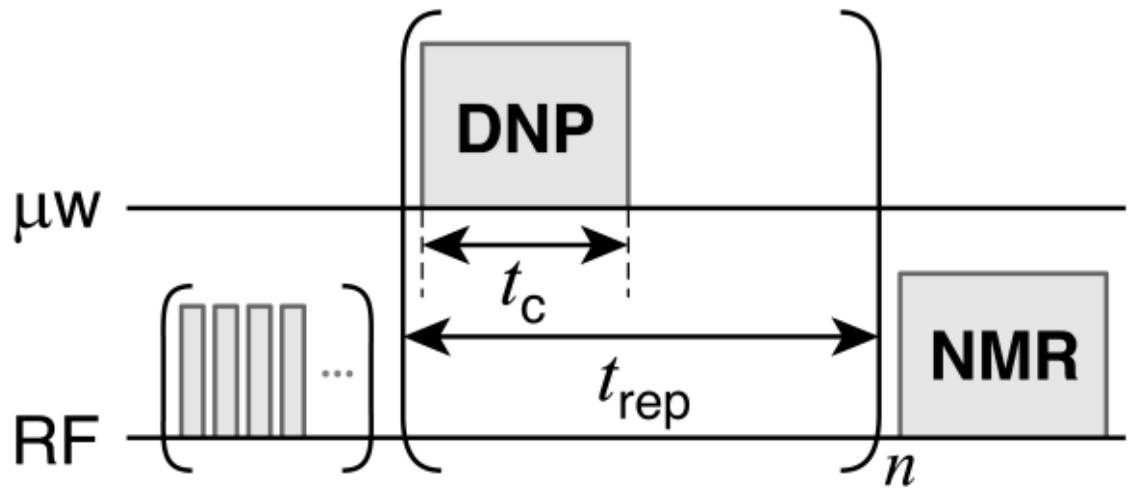

***Figure 3. General scheme of a pulsed DNP experiment.*** *The microwave pulse shape or sequence is applied during the contact time $t_c$ and repeated hundreds to thousands of times with a repetition time $t_{rep}$. This builds up dynamic polarisation of bulk nuclei, which is read out in an NMR experiment. The steady orbit GRAPE algorithm optimises the content of the DNP block to achieve maximum nuclear magnetisation at $n \rightarrow \infty$.*

Two sets of DNP experiments were performed, both following the scheme in Figure 3. In the first set of experiments, linear frequency sweeps were applied with a contact time of 5 µs and a repetition time of 2.5 ms. With the RVA set to 3 dB, the EIKA operated near its full power of 1 3 kW (but still in its linear regime), of which approximately half reaches to the sample. This means that in these experiments, an estimated average microwave power of 1.3 W is applied to the sample. Enhancements of $^1$H NMR signals were observed as a function of microwave resonance offset, resulting in frequency profiles that are characteristic of a DNP waveform. To test and sharpen simulation of pulsed DNP in the steady state[47], the effects of amplitude shaping were explored.

In the second set, the performance of waveforms optimised with steady-orbit GRAPE was compared to XiX[60] and ASE[56]. Contact times were set to 360 ns and repetition times to 167 µs. To steer clear of sample heating due to microwave absorption, the RVA setting was increased to 7.5 dB. An estimated average microwave power of 0.5 W is now applied to the sample. Because freeze-thaw cycles can affect the glassy matrix and DNP performance, the set of benchmarking experiments was completed without interrupting the cooling.

The initial assessment of DNP performance used the intensity of the $^1$H NMR signal after 60 seconds of repeated application of the microwave pulse sequence ( $I_{\text{on}}$ , a representative spectrum is shown in Figure S3a). For each sequence or GRAPE waveform, microwave resonance offset was adjusted to maximise this intensity. To determine the DNP enhancement factor $\varepsilon_{\text{B}}$ relative to the Boltzmann equilibrium magnetisation[85]

$$\varepsilon_{\text{B}} = \frac{I_{\text{on}}}{I_{\text{off}}} \cdot \frac{1 - e^{-d_{\text{off}}^{\text{R}}/T_{1\text{n}}}}{1 - e^{-d_{\text{on}}^{\text{R}}/T_{\text{bu}}}} \tag{14}$$

$^1$H NMR signal without microwave irradiation ( $I_{\text{off}}$ , see Figure S3b for the spectrum), nuclear longitudinal relaxation time $T_{1\text{n}}$ , and the build-up time of the nuclear magnetisation $T_{\text{bu}}$ were also measured. In Eq (14), $d_{\text{on}}^{\text{R}}$ and $d_{\text{off}}^{\text{R}}$ are pulse sequence recycle delays. For $T_{1\text{n}}$ measurements, $^1$H magnetisation was first enhanced with pulsed DNP. Its decay was found to be monoexponential with a characteristic time of 109 s (Figure S3c). To measure $T_{\text{bu}}$ , the time of repeated application of a DNP pulse sequence prior to the $^1$H NMR measurement was varied. Final quantitative assessment of DNP performance was based on the practical sensitivity gain[86]

$$G = \left|\varepsilon_{\text{B}}\right| \sqrt{T_{1\text{n}}/T_{\text{bu}}} \tag{15}$$

which accounts for the fact that a more efficient DNP pulse shape yields a shorter $T_{\text{bu}}$ , which is an advantage if an NMR measurement requires signal averaging.

## 5. Parameter ensembles and instrumental distortions

Control sequence distortions by real-life hardware are unavoidable and sometimes severe – care must be taken to respect instrumental filter functions.[77] HiPER also has a broad distribution of the microwave control field within the sample. Additionally, the sample itself has static orientation and relative spin position disorder[47]. Parameter distribution handling in GRAPE is straightforward: a parallel loop and a weighted sum over the resulting fidelities / gradients / Hessians.[29] Here we focus on the more difficult filter function characterisation aspect.

The control sequence seen by the sample differs from the ideal waveform uploaded into the arbitrary waveform generator. In the linear time-invariant approximation, the filter function of the instrument may be obtained, for example, by measuring (using HiPER's own quadrature detection) the step responses within the XiX sequence. The filter kernel $h(t)$ is then obtained from the definition of causal linear response transformation:

$$y(t) = \int_0^\infty h(\tau)x(t-\tau)d\tau + \eta(t) \quad \Rightarrow \quad \mathbf{y} = \mathbf{Xh} + \boldsymbol{\eta} \tag{16}$$

where $x(t)$ is the input that starts at time zero and has a long enough blank tail for the ringing to fade in the output $y(t)$; and $\eta(t)$ is weak additive noise. In a discrete representation sampled on a uniform time grid, $\mathbf{y}$ and $\mathbf{h}$ become column vectors and $\mathbf{X}$ is then a Toeplitz matrix constructed from the elements of the discretised input vector[87]. A standard method for extracting the kernel is Tikhonov regularisation[88]:

$$\mathbf{h}_\lambda = \left(\mathbf{X}^\dagger \mathbf{X} + \lambda \mathbf{1}\right)^{-1} \mathbf{X}^\dagger \mathbf{y} \tag{17}$$

with the coefficient $\lambda$ chosen using the L-curve criterion[89]. *Matlab* functions performing these operations were implemented and added to *Spinach*. The extracted filter function and its representation in the frequency domain are shown in Figure S4. We do not actually use this method in practice for reasons discussed below, but Figure S4 is nonetheless important because it confirms the applicability of the linear time-invariant system approximation.

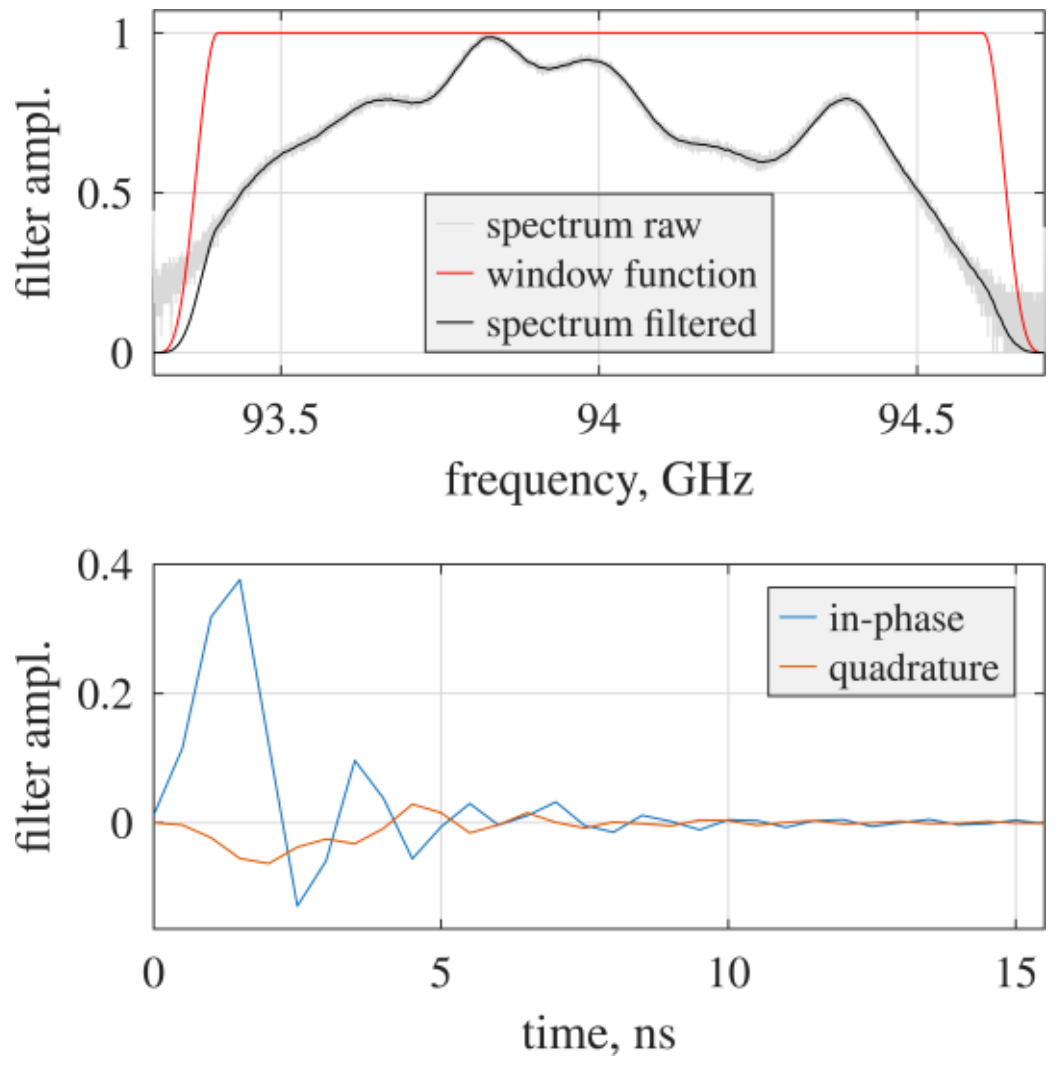

***Figure 4. Waveform distortions induced in HiPER's excitation branch.*** *(Top) Transmission spectrum and (bottom) filter function kernel (in-phase and quadrature components). The indicated window function was applied to the transmission spectrum before extraction of the filter function by causal cepstrum folding.*

The problem with the above method is that the resulting filter kernel also includes significant distortions introduced in the *detection* branch of HiPER, which are not relevant for microwave *excitation*. We therefore decided instead to observe the EIKA output power as a function of microwave frequency, using a directional coupler (blue in Figure 2). The resulting transmission spectrum (Figure 4, top panel) includes only distortions due to synthesis, up-conversion, and, most prominently, amplification of the microwaves; it is dominated by the response of the amplifier. Since HiPER uses a non-resonant sample holder, this is what the electron spins are seeing: excitation of higher-order modes is negligible in the waveguide pipe. The filter kernel (Figure 4, bottom panel) was therefore taken to be the minimum-phase kernel with the measured amplitude spectrum, obtained by the real cepstrum (Kramers-Kronig) construction. The kernel was used to create a differentiable filter function handle for the optimal control module of *Spinach*, enabling the design of HiPER-adapted DNP waveforms as described in our recent paper dealing with instrumental distortions[77]. An important matter is that the transparency window in Figure 4 is physically immutable; this dictates the reference frequency of rotating frame transformations used by simulations. Frequency agility of microwave pulses relative to this frequency must therefore be implemented by varying the phase profile of the pulse waveform.

## 6. Simulation of pulsed DNP in the steady state

The repeated application of a DNP pulse sequence (Figure 3) serves to pump the electron spin polarization from the polarizing agents into the bulk of the sample. Eventually, the complete spin system reaches a steady state in which the nuclear spins are hyperpolarized. For accurate design of a DNP pulse sequence, the effects of its repeated application must be modelled. Simulation of pulsed DNP in the steady state accomplishes this, at minimal computational cost[47]. A variation of the Newton-Raphson root-finding algorithm that solves $\mathbf{P}\boldsymbol{\rho}_{\infty} = \boldsymbol{\rho}_{\infty}$ for $\boldsymbol{\rho}_{\infty}$ is particularly efficient. The propagator $\mathbf{P}$ should cover the evolution of the spin system during the DNP sequence and during the microwave-free waiting time. $\boldsymbol{\rho}_{\infty}$ is then the stroboscopic steady state at the end of the repetition time.

Steady-state simulations were previously tested against experiments, at 0.34 T (15 MHz, 9.7 GHz) for NOVEL,[50] and at 1.2 T (51 MHz, 34 GHz) and 3.35 T for three modulated sequences: TOP[59], XiX[60], and TPPM[61]. We found that trityl radicals in a frozen glassy matrix may be modelled as dipolar-coupled electron-proton spin pairs, with ensemble averaging over a 3.5-20 Å distribution of distances[90]. Anisotropies of the dipolar coupling and the electron Zeeman interaction are accurately handled by finite spherical grid averaging[91]. Longitudinal and transverse relaxation rates of electrons and protons should be set to the correct orders of magnitude, based on experimental information if available.[47] Longitudinal relaxation of protons in close proximity to the radical is distance and orientation dependent and implemented as such.[47,92] Compared to single-shot simulations, steady-state solvers improve two aspects. Firstly, DNP dependence on the microwave resonance offset, the contact time, and the microwave $B_1$ field are more accurately reproduced. Secondly, they make it possible to predict the dependence of DNP efficiency on the repetition time and therefore on the average microwave power.[47]

Figure 5A shows experimental frequency profiles, obtained at 3.35 T, with three amplitude-shaped frequency sweeps. Microwave amplitude is kept constant (blue) or tapered off at the start and the end with a quarter sine wave (green) or with a power of a sine bell (orange, see Figure S5 for the waveforms and the equations). Two types of solid effect DNP are seen. On resonance (94.000 GHz), the $^1$H NMR signal is enhanced via the integrated solid effect (ISE)[53,54,93] in a sequence of three events: an adiabatic passage through the double-quantum condition ($\Omega / 2\pi$ = 143 MHz) followed by an adiabatic inversion of the electron spin magnetisation, followed by a second adiabatic passage through the zero-quantum condition ($\Omega / 2\pi$ = –143 MHz). Simulations in Figure S6 show the evolution of the longitudinal proton magnetisation $\langle \mathbf{I}_z \rangle$ under the ISE. Only at the start and the end of the sweep is magnetisation transferred. During the long middle part, which includes the adiabatic inversion, $\langle \mathbf{I}_z \rangle$ remains approximately constant. This explains why the tapering of the pulse amplitude adversely affects (quarter sine) or completely removes (sine bell power) the ISE: the adiabatic passages are compromised. Away from resonance, the stretched solid effect (SSE) is active[55,57]. It produces positive DNP (stronger) near the DQ matching condition and negative DNP (weaker) near the ZQ condition. Near the DQ condition, pulse amplitude tapering now has a positive effect: DNP enhancements are slightly above those obtained with the constant-amplitude sweep.

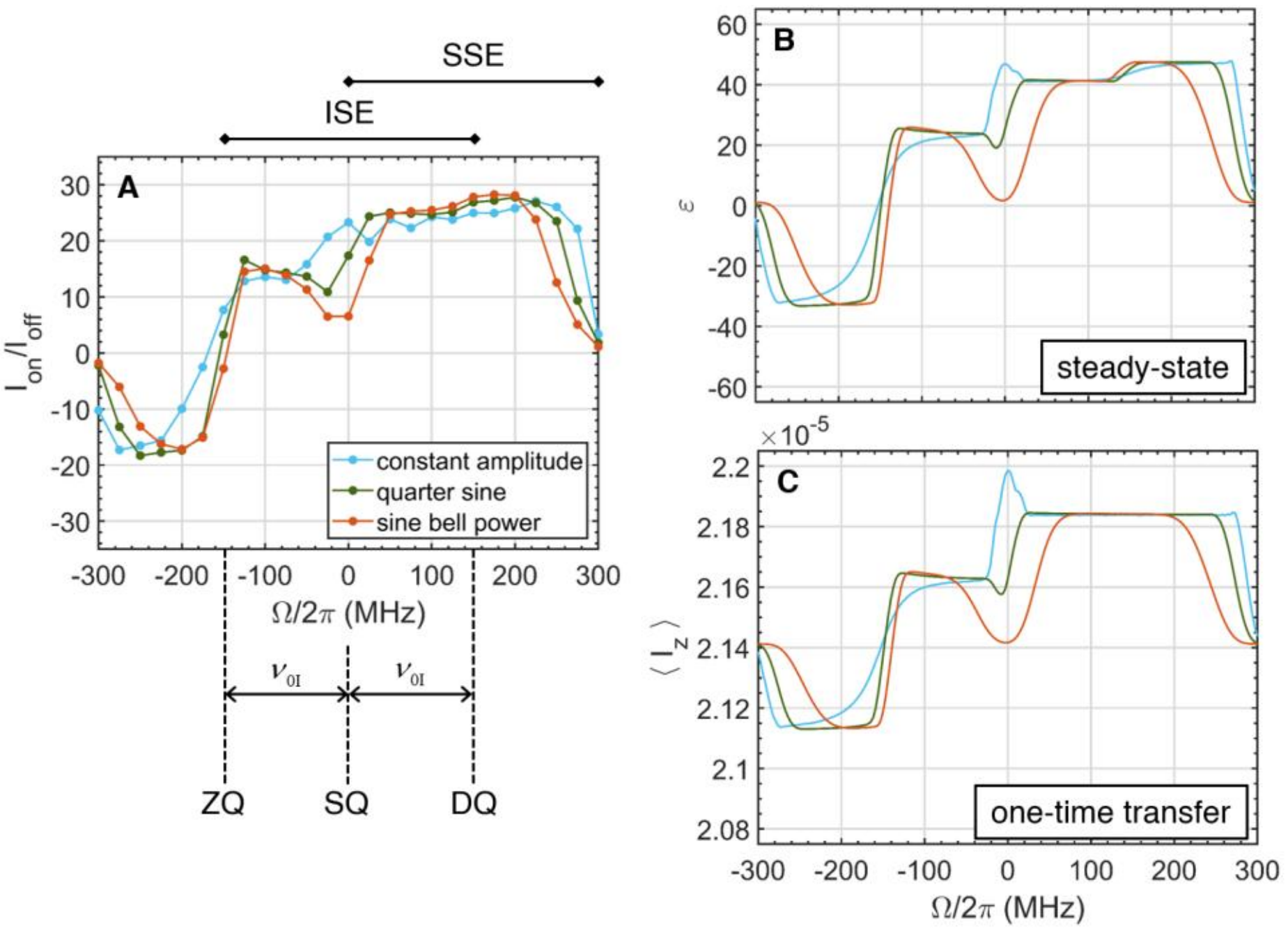

*Figure 5. **Pulsed DNP with shaped frequency sweeps at 3.35 T.** The centre frequency of a linear 300 MHz sweep is varied, relative to 94.000 GHz, and, at each offset, the $^1$H NMR signal intensity is observed. (A) Experiments, along with an illustration of the zero-, single-, and double quantum transitions and the frequency sweeps that give rise to the integrated and the stretched solid effect (ISE and SSE), (B) steady-state simulations, and (C) one-time transfer simulations. The amplitude modulation of the linear frequency sweeps is described in the text and visualised in Figure S5. The RVA was set to 3 dB, the contact time is 5 μs, and the repetition time is 2.5 ms.*

Figures 5B and 5C show steady-state and single-shot DNP simulations of the same frequency profiles. We model the trityl glass as before[47] and here also account for the severe inhomogeneity of the microwave $B_1$ field in HiPER sample holder (Figure S7)[84] using Gauss-Legendre quadrature over a uniform distribution of electron Rabi frequencies between 8 and 42 MHz. We also account for the HiPER instrument filter function (Figure 4). Agreement between steady state simulations and experimental data is excellent. Effects of amplitude shaping are reproduced and so is the asymmetry of the enhancements above and below resonance; it arises from the dip in amplification by the EIKA between 94.15 and 94.25 GHz (Figure 4). Unlike single-shot simulations, the steady-state simulations correctly predict an increase of the SSE enhancements above an offset of 120 MHz and, for the sweep of constant amplitude, a similar maximum efficiency of ISE and SSE. The steady-state simulations are more accurate because they capture the consequences of repeated application of the pulse sequence (Figure 3). Analysis of individual shots reveals that the transfer efficiency of ISE starts out high, but decreases significantly with every DNP cycle. The SSE starts out modestly, but maintains its efficiency better and eventually overtakes the ISE. The reason is basic spin dynamics: magnetisation vectors of spin packets with different Larmor frequencies fan out more readily under on-resonance excitation than under excitation far from resonance. Simulations with different repetition times and pulse lengths (Figure S8) suggest that this effect limits the maximum enhancement that can be obtained with ISE.

## 7. DNP pulse waveform optimisation

Examples of stroboscopic steady state GRAPE DNP pulse sequence optimisations are included with the optimal control example set of *Spinach* 3.0 and later versions (https://github.com/Ilya-Kuprov/Spinach)[67]. We use the thermal equilibrium as the starting state:

$$\boldsymbol{\rho}_0 = \exp\left(-\mathbf{H}/kT\right)/\mathrm{Tr}\left[\exp\left(-\mathbf{H}/kT\right)\right] \tag{18}$$

and seek to achieve maximum longitudinal nuclear magnetisation by setting $\boldsymbol{\sigma} = \mathbf{I}_\mathrm{Z}$ in Eq (11) so that the steady orbit GRAPE algorithm maximises its expectation value with respect to the phase profile of a constant-amplitude pulse.

The drift Hamiltonian is generated automatically by *Spinach*; it includes, at each spin system orientation, anisotropic electron and nuclear Zeeman interactions and the dipolar part of the electron-nuclear hyperfine coupling:

$$\mathbf{H}_\mathrm{D} = \mathbf{S}\cdot\mathbf{Z}_\mathrm{E}\cdot\mathbf{B} + \mathbf{I}\cdot\mathbf{Z}_\mathrm{N}\cdot\mathbf{B} + \mathbf{S}\cdot\mathbf{D}\cdot\mathbf{I} \tag{19}$$

where $\mathbf{B}$ is the static external magnetic field vector, $\mathbf{S}$ is a vector of electron spin projection operators, $\mathbf{I}$ is a vector of nuclear spin projection operators, $\mathbf{Z}_\mathrm{E,N}$ are electron and nuclear Zeeman interaction tensors and $\mathbf{D}$ is the dipolar part of the hyperfine coupling tensor. Appropriate secular and pseudosecular terms[94] are retained in the electron rotating frame. The control Hamiltonian contains the electron microwave irradiation operators, $\mathbf{S}_\mathrm{X}$ and $\mathbf{S}_\mathrm{Y}$, in the rotating frame matched to the chosen reference electron Larmor frequency:

$$\mathbf{H}_\mathrm{C} = \omega_1\left(\mathbf{S}_\mathrm{X}\cos\varphi + \mathbf{S}_\mathrm{Y}\sin\varphi\right) \tag{20}$$

where $\omega_1$ is the electron Rabi frequency and the phase $\varphi$ is varied as a function of time. The filter function from Figure 4 was taken into account as described in our paper on that subject[77].

Time slice duration must be chosen to accommodate three requirements: (a) small electron magnetisation turning angles, ideally below 45° per slice; (b) compatibility with the AWG and the microwave beam delivery system of the instrument; (c) sufficient time resolution to ensure zero-, single-, and double-quantum transitions are within the reachable frequency window. In our case, $\Delta t$ = 0.5 ns achieves all three objectives. The pulse was 360 ns (720 slices) long with a 10 ns (20 slices) ringdown period; the inter-pulse delay was 167 µs. GRAPE optimisation was performed for an ensemble including spin system orientations (800-point REPULSION grid[95]), microwave resonance offset uncertainty (4 MHz range around the intended frequency, sampled in five points, uniformly distributed), and HiPER's $B_1$ inhomogeneity (5 to 25 MHz in 20 points, uniformly distributed), giving a total of 80,000 distinct spin systems. *Spinach* kernel has been upgraded to run in parallel over such ensembles.[29] The optimisations were run, using the *Distributed Computing Toolbox* of *Matlab R2026b*, on a system with two AMD EPYC 9755 128-core CPUs; they took several hours each.

Unusually, some optimal pulses turned out to be interpretable, those are discussed below. Rationally interpretable optimal control waveforms are exceedingly rare in magnetic resonance – we are aware of only one other case in solid-state NMR spectroscopy.[96,97]

## 8. Rapid integrated solid effect

Consider first a favourable case with a narrow distribution of electron Rabi frequencies (16-24 MHz) and no instrumental distortions. In these conditions, optimisations starting from Gaussian white noise consistently converge to an intricate pattern of frequency jumps shown in Figure 6A. Microwave irradiation starts out near the zero-quantum condition (94.000 GHz + 143 MHz), but after just a few nanoseconds, a frequency jump occurs to the double-quantum transition (94.000 GHz – 143 MHz). This is followed by a series of jumps to resonance (94.000 GHz), to the zero-quantum condition, and back to resonance. At this time (105 ns), microwave irradiation becomes briefly bichromatic, with irradiation near both the double and the zero-quantum transition, before another series of frequency jumps is executed, to resonance, to the zero-quantum condition, to resonance, and back to the double-quantum condition. This complete sequence of events is then repeated (starting at 180 ns), in reverse order.

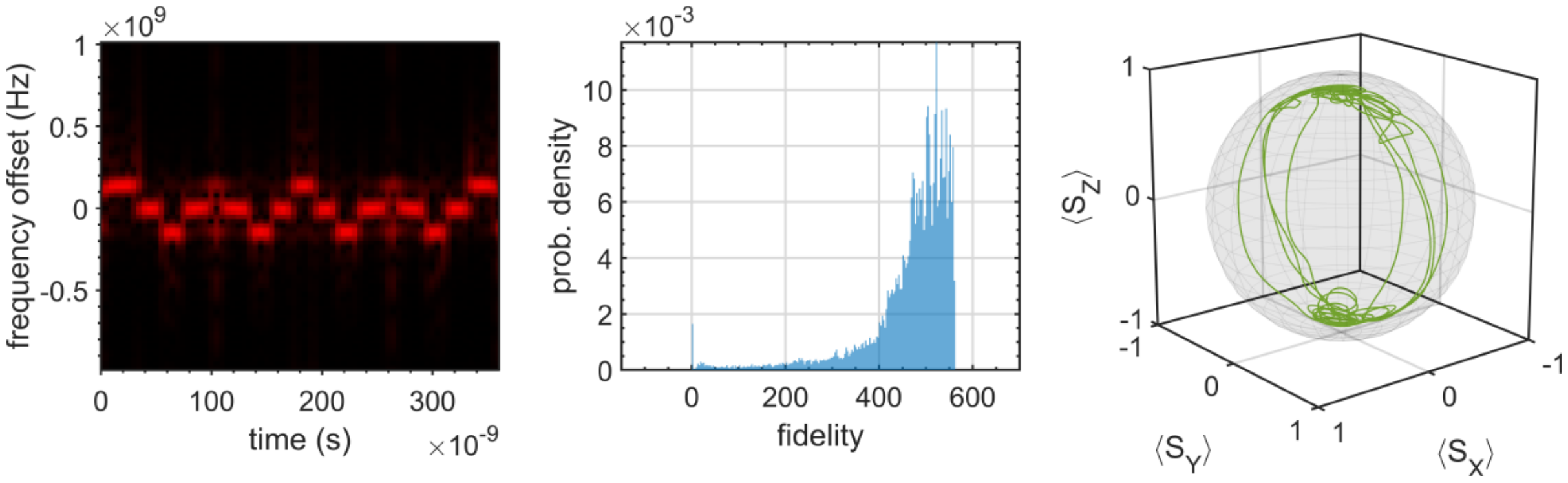

***Figure 6. DNP pulse waveform producing the rapid integrated solid effect (RISE).*** *(left) Microwave pulse spectrogram. (middle) Histogram showing the distribution of the fidelities over 36,000 ensemble instances (grids with 800 orientations, 5 microwave resonance offsets, and 9 electron Rabi frequencies). For convenience, the fidelity is equated to the enhancement factor of the $^1$H polarisation relative to thermal equilibrium. The mean fidelity is 466 and the standard deviation is 98. (right) Steady orbit of the electron spin magnetisation over the course of the 360 ns waveform, plotted for the orientation that gave the highest fidelity at the electron Rabi frequency of 24 MHz and zero resonance offset.*

The fidelity histogram of the optimized waveform (Figure 6B) shows a favourable distribution of enhancement factors: over 400 for most systems in the ensemble. When simulated for realistic conditions of the HiPER instrument (broader distribution of Rabi frequencies, presence of instrumental distortions, electron-nuclear distance distribution – Figure S9), the predicted enhancement factor is around 100, considerably higher than ISE or SSE (Figure 5) even though the average microwave power is 2.8 times lower.

Inspection of the steady orbit of the electron spin magnetisation (Figure 6C) reveals that, by jumping frequencies, the optimal waveform completes the same tasks as the ISE (transfer of magnetisation near the DQ condition, then inversion, then transfer of magnetisation near the ZQ condition) repeatedly and in a much shorter time. This is an adaptation to the narrow ESR spectrum of the trityl radical: DQ, SQ, and ZQ conditions are well separated and irradiation at intermediate frequencies is unnecessary. This is a predictable optimisation – ISE was originally developed to transfer polarisation from the photoexcited triplet state of pentacene, which, due to inter-electron dipolar interactions, has a much broader ESR spectrum.[53] Accordingly, we call the new waveform *rapid integrated solid effect* (RISE).

The periods of on-resonance irradiation (Figure 6A) causing magnetisation inversions last approximately 25 ns. This corresponds to the duration of a 180°-pulse at a Rabi frequency of 20 MHz as well as to the duration of an adiabatic inversion with a $Q_{\mathrm{crit}}$ of 3. The latter is just about sufficient to complete the inversion at a Rabi frequency of 24 MHz, if one assumes a linear frequency sweep over 30 MHz.[98] The resolution of the spectrogram cannot separate one from the other, but a closer look at the steady orbits indicates that the actual RISE waveform is more complex and tailored. Pole-to-pole inversion is complete only for the highest Rabi frequencies in the distribution; other spin packets reach intermediate latitudes in the southern hemisphere, from where they follow predominantly transverse trajectories until the next inversion period. On their way back up, magnetisation vectors regroup. This process is consolidated during the periods of bichromatic and double-quantum irradiation. We think that periodic regrouping ensures synchronous magnetisation transfer within ensembles: the growth of longitudinal $^1$H magnetisation is remarkably gradual and synchronous across orientations (Figure S9).

## 9. Experimental evaluation of optimal waveforms

We now proceed to design and test DNP waveforms that accommodate the full (ensemble plus instrument distortions) complexity associated with a real sample in a real instrument.

The RISE waveform from the preceding section was re-optimised for HiPER's broad electron Rabi frequency distribution (Figure S7) and its distortion filter kernel (Figure 4). The spectrogram undergoes notable changes (Figure 7, top panel): periods of irradiation around the DQ condition have lengthened, while those around the ZQ condition have shortened and have become bichromatic. We attribute this to the instrument filter function – at the ZQ condition, the microwave transparency of HiPER is only 60% (Figure 4). The optimisation responds by staying away from the ZQ condition and transferring magnetisation through the DQ condition as much as possible.

Even when distortions are taken into account, they clearly scramble the pulse sequence and its performance deteriorates. The fidelity histogram (Figure 7, first row) reports a large number of subsystems with little to no enhancement; the mean fidelity is reduced to 301. The steady orbit of the electron spin magnetisation (Figure 7, first row) becomes more intricate and now defies visual interpretation. Our impression is that the optimiser still tries to keep magnetisation vectors together, as in the idealised version of RISE in Section 8, but no longer succeeds, likely due to the broad $B_1$ field distribution. At a Rabi frequency of 5 MHz, a 180° pulse takes 100 ns – repeated inversions that previously regrouped magnetisation vectors are no longer possible.

Experiments with HiPER-adapted RISE at 3.35 T yielded an $I_{\mathrm{on}}/I_{\mathrm{off}}$ of 34.0 (Table 1), respectable performance for a sample volume of 30 μL and an average microwave power of 0.5 W. We observed a lower $^1$H NMR signal intensity ($I_{\mathrm{on}}/I_{\mathrm{off}}$ = 26.1) with XiX, which – with an optimised pulse length of 18 ns – was the only modulated DNP sequence that produced a significant enhancement factor. The second row of Figure 7 shows the spectrogram of XiX along with its fidelity histogram and steady orbit. The orbit shows the electron spin magnetisation vector doing two 2π turns around the effective frequency vector $\boldsymbol{\omega}_{\mathrm{eff}}$ tilted towards –X, followed by two 2π turns around the effective frequency tilted towards +X, and so on. In the fidelity histogram, only a small fraction of the ensemble has enhancement factors close to 600, the majority stays below 50 and some are even negative. A deeper investigation reveals that *g*-tensor anisotropy, although small for the trityl OX063 radical, is the cause of these poorly performing orientations. We conclude that the original XiX sequence, despite its relatively good performance at low peak powers, is less suitable for pulsed DNP at high field.

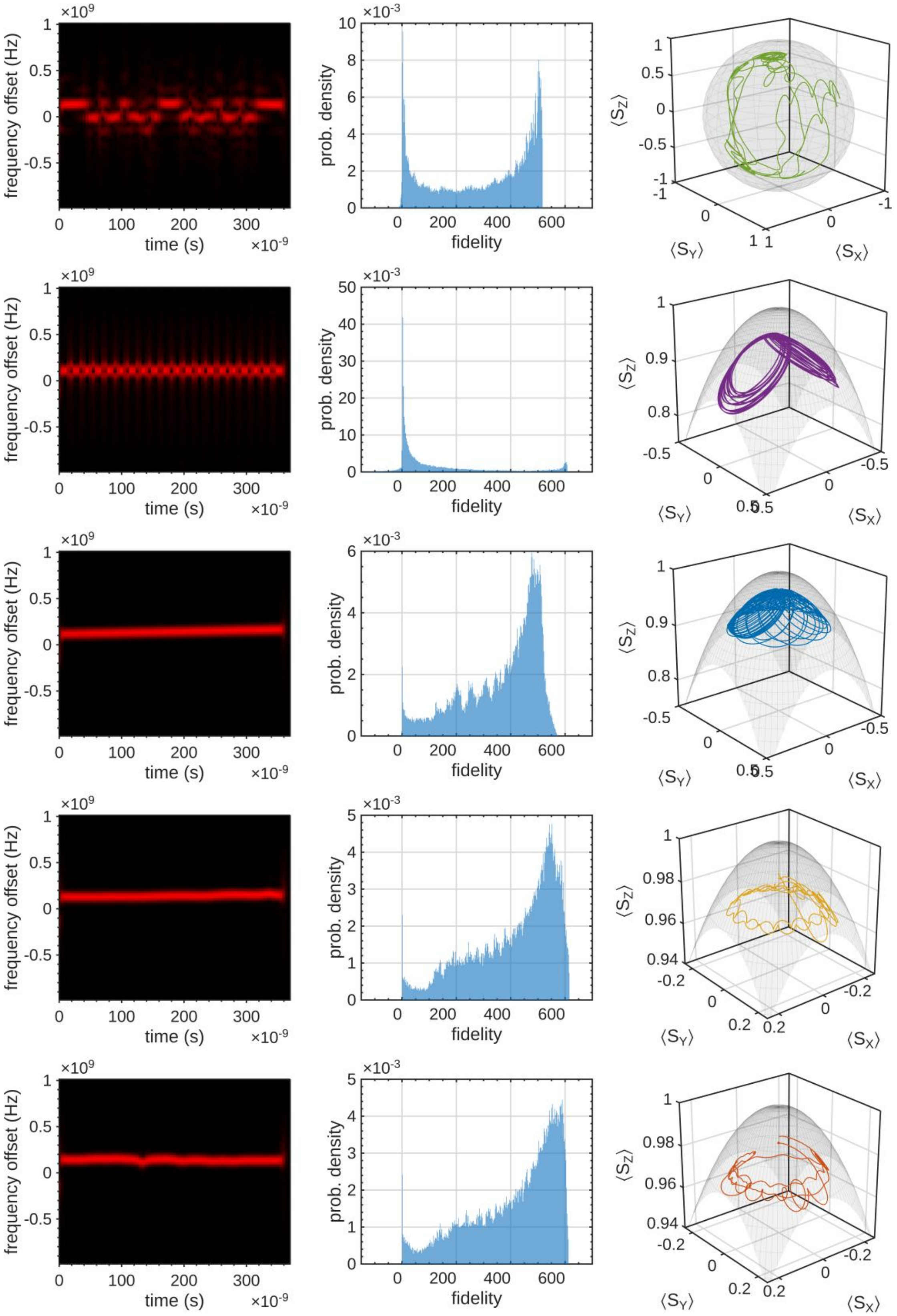

*__Figure 7. HiPER instrument filter function adapted DNP pulse waveforms and their performance.__ From top to bottom: HiPER-adapted RISE, XiX, ASE, EASE, and ODQ. Spectrograms were obtained by taking Fourier transforms of a moving time window. Histograms show the distribution of the fidelities over the 80,000 spin systems (orientation, microwave resonance offset, and electron Rabi frequency ensembles). Steady orbits of the electron spin magnetisation (normalised to thermal equilibrium) are plotted for the orientation that gave the highest fidelity at an electron Rabi frequency of 25 MHz and offsets of 0 MHz (for HiPER-adapted RISE), 110 MHz (for XiX), and 140 MHz (for ASE, EASE, and ODQ). See the text and Table 1 for further specifications.*

Unlike integrated and stretched solid effect pulse sequences, the adiabatic solid effect (ASE) experiment only has one narrow linear frequency sweep through either the double- or the zero-quantum condition. We chose the DQ condition, because our instrument delivers more power there (Figure 4) and optimised the sweep width (Figure S10): 30 MHz (Figure 7, third row) yielded $I_{on}/I_{off}$ of 51.4 (Table 1). The targeted adiabatic passage of ASE thus proves superior to repeated inversions and frequency jumps. The steady orbit (Figure 7, third row) is typical for an adiabatic passage: the magnetisation follows the slowly moving effective field as it precesses around it. A turning point occurs when the microwave frequency becomes equal to the frequency of the rotating frame. An advantage of adiabatic passage over merely fulfilling a matching condition is that orientations that have slower magnetisation transfer (due to weaker dipolar coupling or a frequency mismatch) have a chance to catch up with orientations that have faster transfer. This is apparent when monitoring the evolution of the longitudinal nuclear magnetisation for individual orientations (Figure S11). However, the fidelity histogram (mean enhancement factor of 368 and a standard deviation of 140) suggests that there may be room for improvement in ASE – we therefore attempted optimal control.

***Table 1. Benchmarking of HiPER-adapted waveforms against XiX and ASE.*** *The values in brackets following the enhancement factors predicted by the optimiser are the standard deviations of the fidelity distributions. $I_{off}$ was measured with a recycle delay of 180 s (Figure S3b), $I_{on}$ was measured for each sequence after 60 s of irradiation and following optimisation of the microwave resonance offset (Figure S12). $T_{1n}$ is 109 s (Figure S3c). In terms of practical sensitivity gain, ODQ constitutes an improvement of 5.8% over ASE, with the 95% confidence interval ranging from 0.4% to 11.5%. For the uncertainty analysis, see the Supporting Information.*

| DNP sequence | Simulation | | Experiment | | | |
|---|---|---|---|---|---|---|
| | $\varepsilon_B$, GRAPE | $\varepsilon_B$, incl. $r$ ensemble | $I_{on}/I_{off}$ | $T_{bu}$ (s) | $\varepsilon_B$ | $\varepsilon_B\sqrt{(T_{1n}/T_{bu})}$ |
| RISE | 301 (181) | 97 | 34.0 | - | - | - |
| XiX | 128 (179) | 45 | 26.1 | - | - | - |
| ASE | 368 (140) | 98 | 51.4 | 60.2 | 65.8 | 88.5 |
| EASE | 413 (153) | 109 | 49.4 | - | - | - |
| ODQ | 412 (158) | 112 | 54.7 | 57.1 | 67.9 | 93.7 |

When ASE is used as the initial guess, steady orbit GRAPE succeeds in improving it. In the spectrogram of the optimised waveform (Figure 7, fourth row), which we will call EASE (efficient adiabatic solid effect), not much initially appears to have changed, apart from a few wobbles. However, the fidelity histogram looks better (average enhancement factor increases to 413) and the steady orbit (Figure 7, fourth row) undergoes drastic changes: precession around the slowly moving effective field disappears, resulting in a horizontal orbit with only minor oscillations and loops. Such an orbit, with the transverse components maximised, cannot be generated with pulses of constant amplitude and phase: it can only the product of optimal control. It is an efficient solution because, in the effective Hamiltonian, the double-quantum ($\mathbf{S}_{\pm}\mathbf{I}_{\pm}$) and zero-quantum ($\mathbf{S}_{\pm}\mathbf{I}_{\mp}$) terms are responsible for the transformation of longitudinal electron magnetisation into longitudinal nuclear magnetisation. For the steady orbit of EASE the effective frequency is undefined, but visual inspection indicates that its time-averaged direction is along the Z axis; Palani *et al.* have pointed out that this is a suitable choice for the Z axis of the interaction frame.[46]

The EASE waveform and its steady orbit improve magnetisation transfer in three more ways. First, at the end of the contact time, electron magnetisation is returned to the Z axis. This speeds up the preparation for the next round of magnetisation transfer by reducing the wait for longitudinal relaxation to bring the electron magnetisation back to thermal equilibrium. It is similar to the flip-back pulse at the end of the NOVEL sequence[48]. Second, the orbits are synchronised across the entire electron Rabi frequency distribution, making sure that the microwave irradiation is effective for all spin packets. Third, EASE maintains the adiabatic character of ASE: $\langle \mathbf{I}_z \rangle$ evolution curves in Figure S11 are very similar to those of ASE (still allowing slow orientations to transfer magnetisation), but without unnecessary sidetracks.

Elegant as it may look, EASE does not beat the state of the art (Table 1). This is likely because GRAPE solutions are nearest local optima and their performance depends on the initial guess. To explore this further we ran another optimisation starting from a monochromatic pulse at the double-quantum condition. This resulted in a waveform we call ODQ (optimised double-quantum) that comes out on top in practice as well as in theory (Table 1). The steady orbit of ODQ (Figure 7, bottom row) is similar to that of EASE: it is horizontal, magnetisation is returned to the Z axis at the end of the contact time, orbits are synchronous, and $\langle \mathbf{I}_z \rangle$ evolution curves (Figure S11) are adiabatic. Small differences between the orbits of EASE and ODQ include minor oscillations at the top of the horizontal orbit, initial approach to the frequency of the rotating frame (140 MHz) from the opposite direction, and the presence of multiple turning points. In experiments, ODQ wins on performance with $I_{on}/I_{off}$ = 54.7 (Table 1). When the build-up times of the bulk nuclear magnetisation are taken into account (Figure S13), ODQ yields a 5.8% higher practical sensitivity gain relative to ASE: an achievement indeed, since adiabatic DNP pulse sequences are in some ways theoretically optimal.

## 10. Further development outlook

We offer the following improvements relative to the state of the art in DNP: steady orbit optimal control, instrumental distortion handling, and large ensemble support (suitable for amorphous solids), all implemented in *Spinach*. In experimental testing, we managed to beat the state-of-the-art DNP pulse sequences, but only by a small margin; that requires a discussion.

Optimised RISE waveform is an adaptation of the ISE for narrow-line radicals. In experiments at 3.35 T, RISE could not withstand instrumental distortions, but its method of pulse bandwidth focusing to avoid unproductive frequencies is relevant for polarisation transfer from photo-excited triplet states where ISE is traditionally used. In 2025, Sakamoto *et al.* demonstrated triplet DNP from substituted fullerenes which have, compared to pentacene, a narrow ESR spectrum[99]. We expect RISE type sequences to find applications in that area.

When the electron Rabi frequency is small compared to the nuclear Larmor frequency, targeted irradiation at the DQ or ZQ condition is the most efficient means of transferring polarisation from narrow-line radicals. This is embodied in the narrow frequency sweep of the ASE (compare Figures S8 and S10) which generates higher enhancements at 3.35 Tesla than both HiPER-adapted

RISE and the modulated sequences (Table 1). Frequency sweeps, whether used for adiabatic inversion[100] or magnetisation transfer via adiabatic passage[58], are a mainstay of magnetic resonance because they are robust to real-world distortions, such as inhomogeneous line broadening, distributions of Rabi frequencies, and even transients in amplifier output upon frequency jumps. Nevertheless, optimal control of the steady orbit produces a better performing waveform, ODQ. The horizontal steady orbit of the electron magnetisation in ODQ resembles LOOP (longitudinally optimised with overarching periodicity) DNP sequences[66], in which $\boldsymbol{\omega}_{\text{eff}}(t)$ also aligns closely with the Z axis in the interaction frame. LOOP sequences, however, were designed with the purpose of increasing bandwidth by removing the need for a flip pulse to align equilibrium electron magnetisation with $\boldsymbol{\omega}_{\text{eff}}(t)$ prior to excitation around the single quantum transition[49,62,61]. In ODQ, the horizontal orbits serve to maximise the double- and zero-quantum terms in the effective Hamiltonian for a restricted electron Rabi frequency[101]. Additional synchronisation of orbits over all spin packets and straightening out unnecessary wobbles assures that ODQ uses the available microwave irradiation as efficiently as possible. Returning the electron magnetisation to the Z axis at the end of the contact time prepares the spin system for the next round of the pulse sequence.

Our spin dynamics simulations are sophisticated[47] but still incomplete. The structure of the polarising agent and spin diffusion in the frozen glassy matrix, which play critical roles in the propagation of the magnetisation through the sample, are not explicitly modelled. Electron-electron interactions, suspected to have an effect on DNP pulse sequence performance[61], are not included at all. This may be the reason our simulations underestimate the performance of ASE and overestimate the performance EASE and ODQ (Table 1). Extending spin dynamics simulations in this direction requires special measures because spin systems become exceedingly large. Much progress was recently made using restricted state spaces[102–104] and the Landau-Zener approximation[105]; those methods will likely also work for optimal control. Ensemble averaging over conformations sampled from molecular dynamics simulations is needed to capture the amorphous nature of the frozen glassy matrix. These improvement directions are clear, but the scale of the calculations involved remains daunting.

The standing wave pattern in HiPER's non-resonant sample holder forced us to design DNP pulse sequences that are effective across a broad distribution of electron Rabi frequencies (Figure S7). Since electron Rabi frequencies are also very inhomogeneous in magic angle spinning (MAS) rotors[106], this ability and the resulting sequences are pertinent to the development of pulsed MAS DNP.

To enable benchmarking of DNP pulse sequences on HiPER without interference from sample heating, we restricted the peak microwave power at the sample to 230 W (RVA set to 7.5 dB) and the average power to 0.5 W. It appears that the partial removal of the metal waveguide in the DNP probe impairs heat conduction. In MAS DNP, heating due to microwave irradiation is mitigated by cold bearing and drive gas: in current MAS DNP experiments,[107] gyrotrons routinely provide high-frequency microwaves continuously at over 10 W; a klystron at 263 GHz provides about 5 W.

Recently introduced single crystal diamond rotors will improve thermal conductivity relative to sapphire and zirconia[108] and are expected to offer further relief, thereby allowing more microwave power to be used and higher sensitivity to be obtained.

Limited availability of high-power, high-fidelity microwave sources at frequencies above W-band (75-110 GHz) currently restricts the experimental implementation of pulsed DNP to relatively low magnetic fields. Construction of mm-wave amplifiers is a topic of ongoing engineering effort[109,110] and prototypes are operational in specialised laboratories[111]. Amplification bandwidth is a figure of merit for these devices and depends on the precise machining of the interaction structure. This implies that microwave pulse bandwidth will always be restricted in an instrument-specific way as we experienced with HiPER (Figure 4). The ability to combine steady orbit GRAPE with instrumental distortion handling is therefore valuable. This goes *a fortiori* for the design of pulse sequences for DNP with biradicals[112] and metal ions[113] that have broad and intricate ESR spectra. Pulsed excitation is likely to offer advantages over continuous-wave irradiation; we expect optimal control to be decisive in DNP pulse sequence design.

## 11. Conclusions

We have introduced an optimal control theory and a numerical algorithm that drives quantum systems towards user-specified steady orbits when a control sequence is applied repeatedly. The orbits are constrained by a time-ordered sequence of projectors into intermediate states or subspaces: the fidelity incurs a penalty whenever the system does not respect the specified projectors at the specified times. For an asymptotically attracting steady orbit to exist, the system model must be dissipative; our algorithm then maximises the specified measures of fidelity at the specified points in the steady orbit. The method is different from the prior art (Floquet-Lindblad methods, effective Hamiltonian theories, *etc.*) which has so far focused on average Hamiltonians and stroboscopic steady states: here, the entire limit cycle is the target.

At the implementation level, the algorithm is a modification of Liouville-space GRAPE[29] and inherits its numerical efficiency. It is available, along with the examples that generate Figures 4-7, in versions 3.0 and later of *Spinach* library (https://github.com/IlyaKuprov/Spinach)[67]. The utility of the method is illustrated here for a DNP experiment in a spin system, but the *Spinach* module is written so as to be applicable to any dissipative dynamics that is representable by an exponential action of an evolution generator on a state vector.

## Associated content

**Supporting Information** Detailed HiPER scheme; ESR spectrum and relaxation measurements; $^1$H NMR signals and relaxation; HiPER filter function via quadrature detection; amplitude shaped linear frequency sweeps; simulated contact curves with ISE; HiPER's electron Rabi frequency distribution; pulse lengths and repetition times of ISE and SSE; RISE contact curves and frequency profile; sweep widths, pulse lengths, and repetition times of ASE; HiPER-adapted waveforms contact curves and frequency profiles; experimental offset optimisation and build-up curves; uncertainty analysis.

**Data availability** Numerical algorithms and examples of their application are available in versions 3.0 and later of *Spinach* library (https://github.com/IlyaKuprov/Spinach). All experimental data generated in this study will be deposited on Zenodo after peer-reviewed publication.

## Acknowledgements

This work was funded by the Deutsche Forschungsgemeinschaft through SFB 1527 (Project No. 454252029) and by EPSRC (EP/Y035267/1). The work at the Weizmann Institute was supported by research grants from the Center for New Scientists at the Weizmann Institute of Science and from Anton Rabie, the Stanley and Tanya Rossby Endowment Fund, Danielle Bitton & Raphy Benbaron, and two benefactors who have chosen to remain anonymous. It was also made possible in part by the generosity of the Harold Perlman Family. Computational work was carried out using Weizmann Faculty of Chemistry high-performance computing facility CHEMFARM, which is supported in part by the Ben May Center for Chemical Theory and Computation. Matlab technical support from the awesome engineering team at MathWorks is also acknowledged.

SUPPLEMENTARY INFORMATION

# Quantum optimal control of steady orbits

Shebha Anandhi Jegadeesan[1,#], Yujie Zhao[2,#], Robert Hunter[2], Hassane El Mkami[2], Maximilian Keitel[3], Callum Musselwhite[3], Graham Smith[2,*], Guinevere Mathies[1,*], Ilya Kuprov[4,*]

[1]*Department of Chemistry, University of Konstanz, Konstanz, Germany.*

[2]*School of Physics and Astronomy, University of St Andrews, St Andrews, United Kingdom.*

[3]*School of Chemistry and Chemical Engineering, University of Southampton, Southampton, United Kingdom.*

[4]*Department of Chemical and Biological Physics, Weizmann Institute of Science, Rehovot, Israel.*

## Contents

*Email: *gms@st-andrews.ac.uk*
*guinevere.mathies@uni-konstanz.de*
*ilya.kuprov@weizmann.ac.il*

[#]These authors contributed equally.

## S1. HiPER microwave bridge

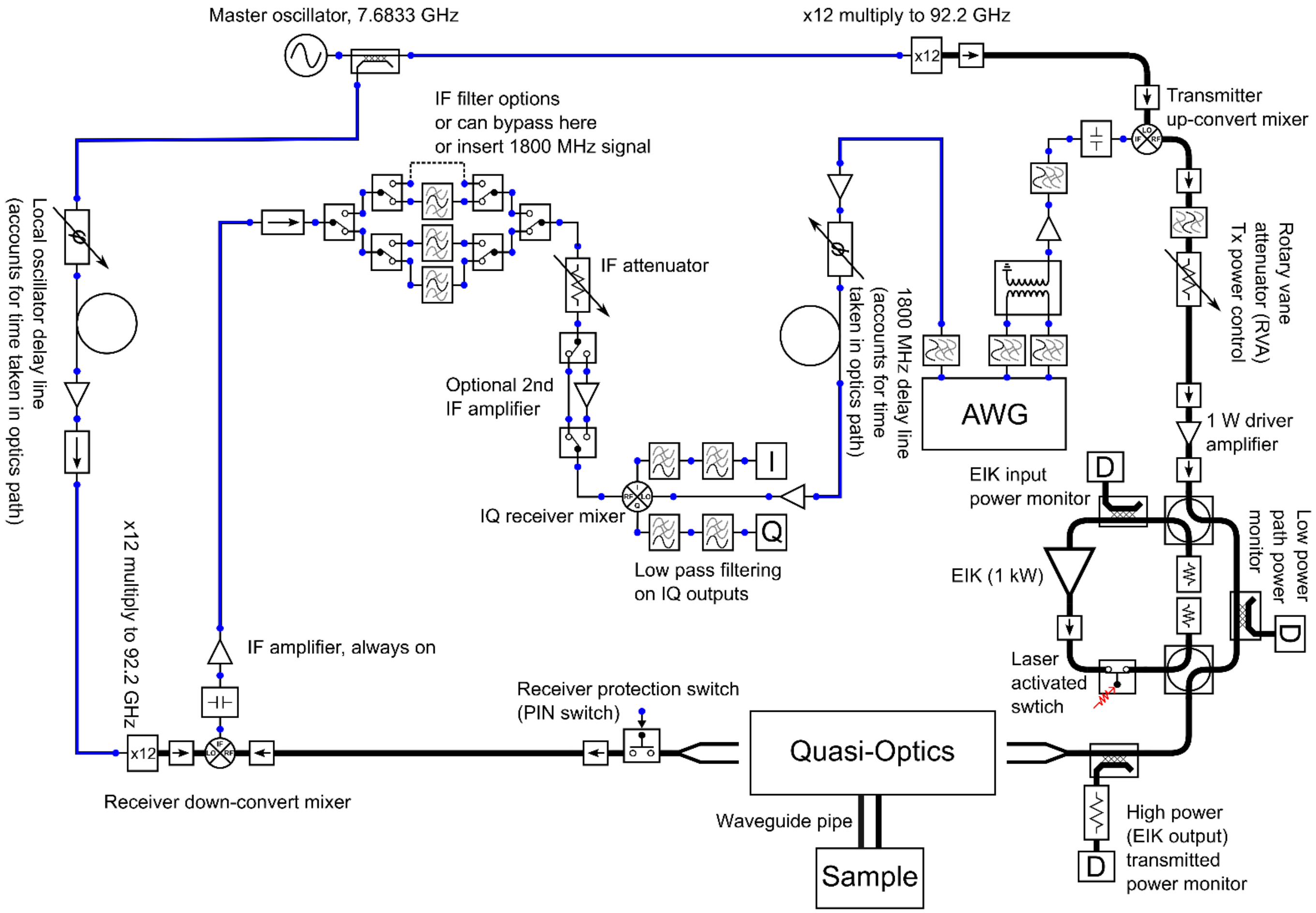

***Figure S1:*** *Detailed diagram of the HiPER microwave bridge. For the purpose of measuring the filter function with HiPER's quadrature detection, a roof mirror was placed at the bottom of the waveguide pipe. A low-noise amplifier at 94 GHz normally improves the signal-to-noise ratio of the EPR detection by a factor of 1.4, but it was not used in the experiments reported in this work.*

The output of a 7.6833 GHz dielectric resonator oscillator (DRO) is split and fed into two ×12 multipliers. One of the resulting 92.2 GHz signals serves as the local oscillator (LO) in the up-conversion by the transmitter mixer, the other as the local oscillator in the down-conversion by the receiver mixer.

A Keysight M8190A (12-bit, 12 GSa/s) arbitrary waveform generator (AWG) creates the desired waveform using digital IQ modulation of a numerically controlled oscillator at 1.8 GHz ± 0.5 GHz with a time resolution of 0.0833 ns. This signal serves as the intermediate frequency (IF) in the transmitter mixer, enabling amplitude and phase control at 94 GHz over a 1 GHz bandwidth. High-performance filters are used to eliminate the local oscillator frequency and unwanted sidebands.

An extended interaction klystron amplifier (EIKA) amplifies the 94 GHz microwaves by 45 dB. A rotary vane attenuator (RVA) that precedes a 1 W solid-state amplifier and the EIKA, is used to control peak output power, up to a maximum of 1.3 kW. The EIKA can operate at a repetition rate of up to 80 kHz, with a maximum continuous pulse length of 12 μs, and a maximum average power of 130 W. In practice, average power is restricted by software to about 3 W to protect the waveguide isolator placed at the EIKA output. A laser-activated switch can be used for gating. If required to remove dark noise following the final pulse or for low-power excitation, the EIKA may be left out of the transmitter arm.

Quasi-optics act as a circulator, routing the transmitted 94 GHz signal to the sample inside the magnet and the ESR signal from the sample to the receiver, as detailed in the original HiPER publication.[1] The main excitation pulse is reflected back towards the transmitter, but is absorbed by high-performance quasi-optical isolators that provide over 90 dB isolation between sample holder and transmitter with < 1dB of insertion loss. Detection of the EPR signal at linear polarisation orthogonal to the polarisation of the excitation (*via* circularly polarized magnetic resonance transitions) typically provides 40 dB broadband isolation and up to 80 dB narrowband isolation between the transmitter and the receiver. Isolation between the sample and the receiver is 70 dB. To protect the receiver, a PIN switch is automatically closed when the EIKA is firing.

ESR signals at 94 GHz are mixed down back to 1.8 GHz at the receiver mixer, amplified, and filtered. Using the second output from the AWG, signals are coherently down-converted to baseband in a quadrature detection scheme. An analogue-to-digital convertor (ADC) card inside the control computer, capable of fast averaging with 1 GHz analogue bandwidth, is used to record the signal. With the HiPER instrument, ESR signals can be detected with a time resolution of less than a nanosecond and very short dead time following excitation with a kilowatt pulse.

## S2. ESR measurements at 3.35 T (94 GHz)

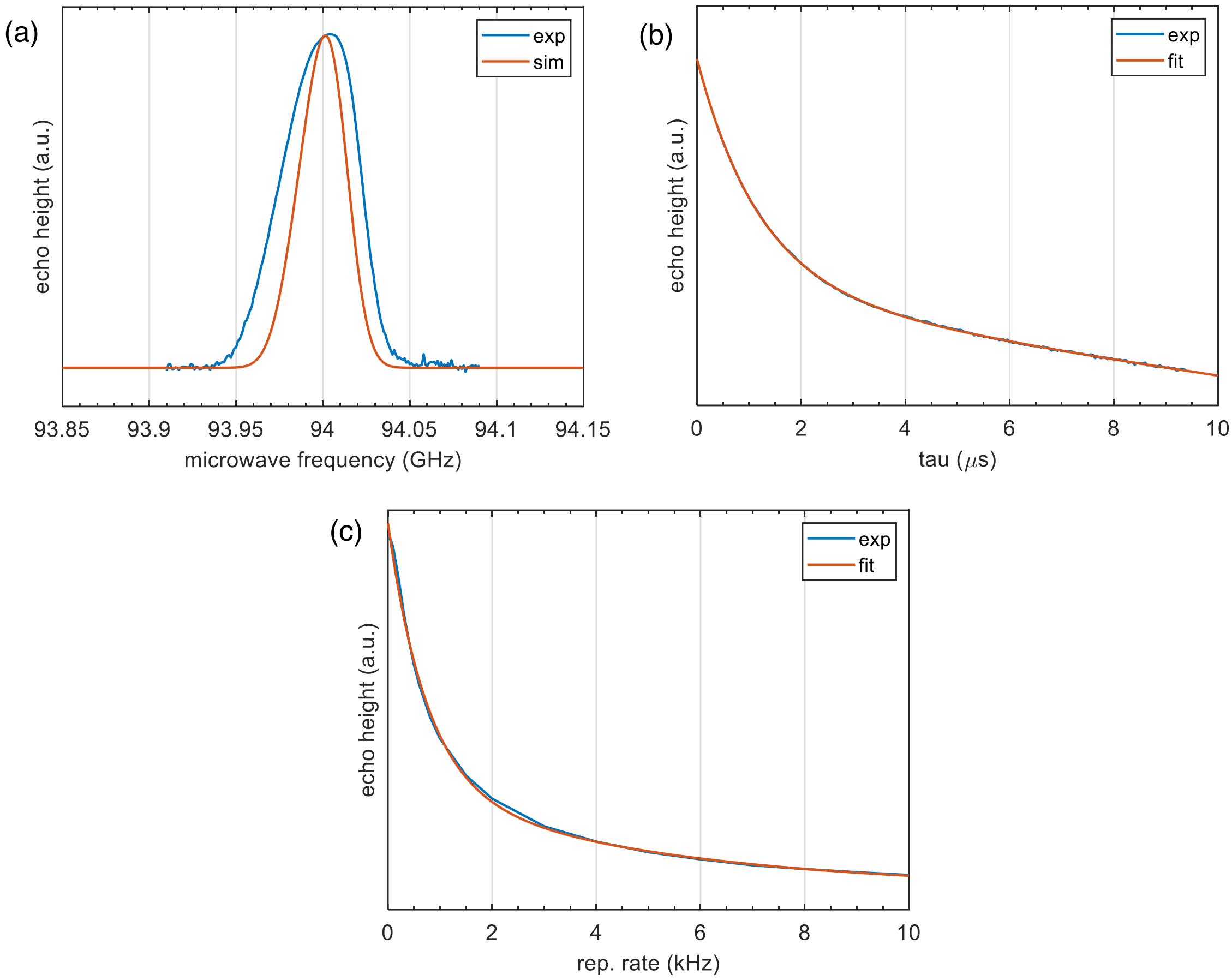

***Figure S2:*** *ESR characterisation of 4.6 mM (verified by UV-Vis spectrophotometry) trityl OX063 in "DNP juice" at 80 K. (a) Echo-detected frequency scan recorded with a soft-pulse echo sequence (150-450-150-450 ns). The simulation was done using EasySpin[2] at a magnetic field of 3.353 T. g-tensor anisotropy ($g_x$ = $g_y$ = 2.003193, $g_z$ = 2.002583) and g-strain (a Gaussian distribution with a width of 0.00035 around the specified g eigenvalues) are those reported by Lumata et al.[3] Isotropic Gaussian broadening of 15 MHz was added on top. The full width at half-maximum (FWHM) of the simulated spectrum is 32 MHz. The FWHM of the experimental spectrum is 52 MHz. (b) Decay of a soft-pulse echo (100-τ-100-τ ns). Fit with a biexponential decay yields 2.4 μs and 77 μs as characteristic times. (c) Electron longitudinal relaxation time $T_{1e}$ was estimated by monitoring the effect of the repetition rate on the echo intensity. Fit with a biexponential decay yields 1.4 ms and 0.21 ms as characteristic times.*

## S3. $^1$H NMR measurements at 3.35 T (143 MHz)

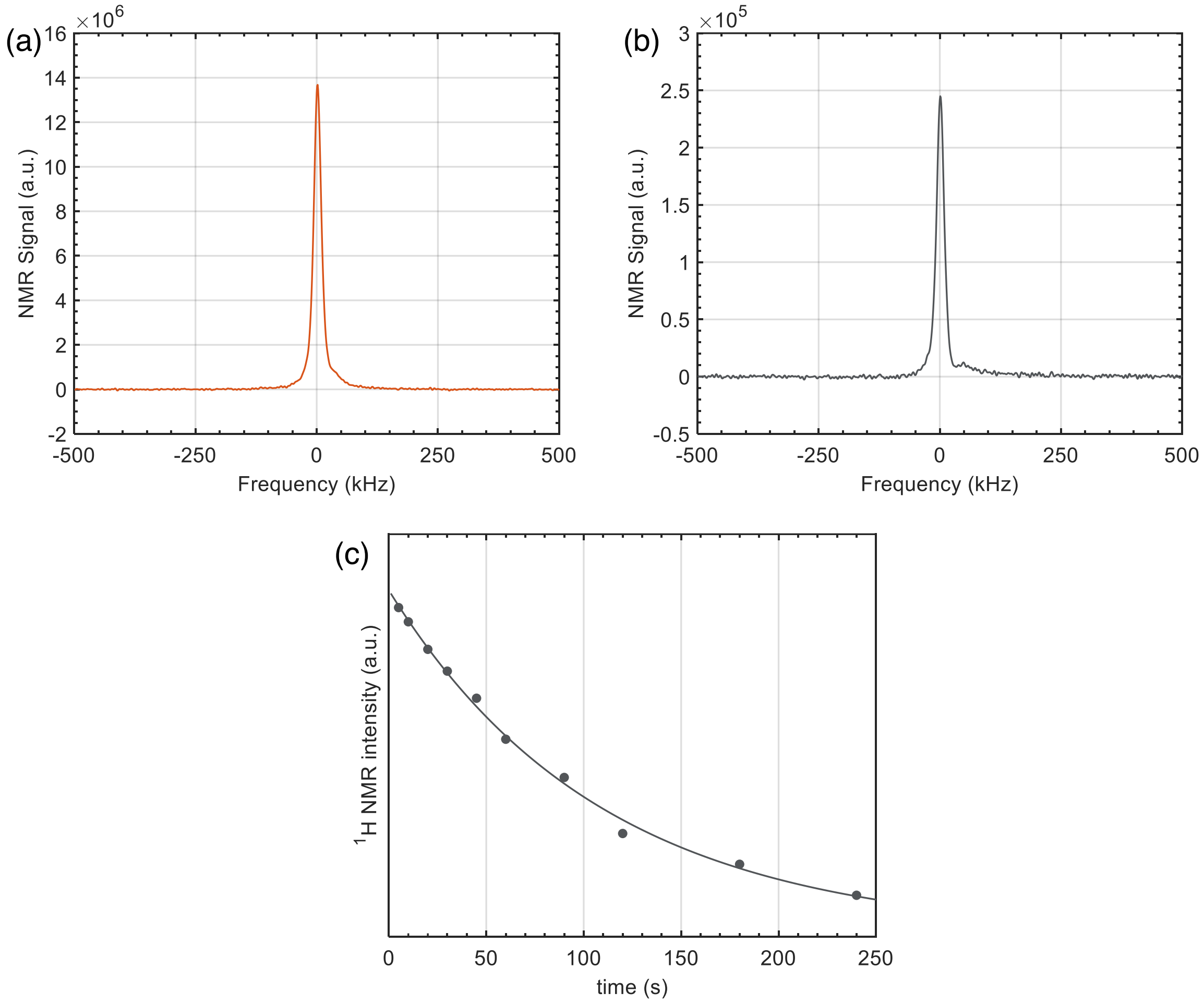

***Figure S3:*** *$^1$H NMR of 4.6 mM trityl OX063 in "DNP juice" at 80 K. (a) Single acquisition after 60 s irradiation with the ODQ DNP pulse sequence. Integral (–150 to 150 kHz) = 3.51×10$^8$ a.u. (b) Without DNP, average of 260 acquisitions, recycle delay 180 s. Integral (–150 to 150 kHz) = 6.29×10$^6$ a.u. (c) Decay of the $^1$H NMR signal to thermal equilibrium following enhancement by pulsed DNP. Fit with a mono-exponential decay yields a $T_{1n}$ of 109 s, with a standard error of 9.9 s.*

# S4. Filter function using HiPER's own quadrature detection

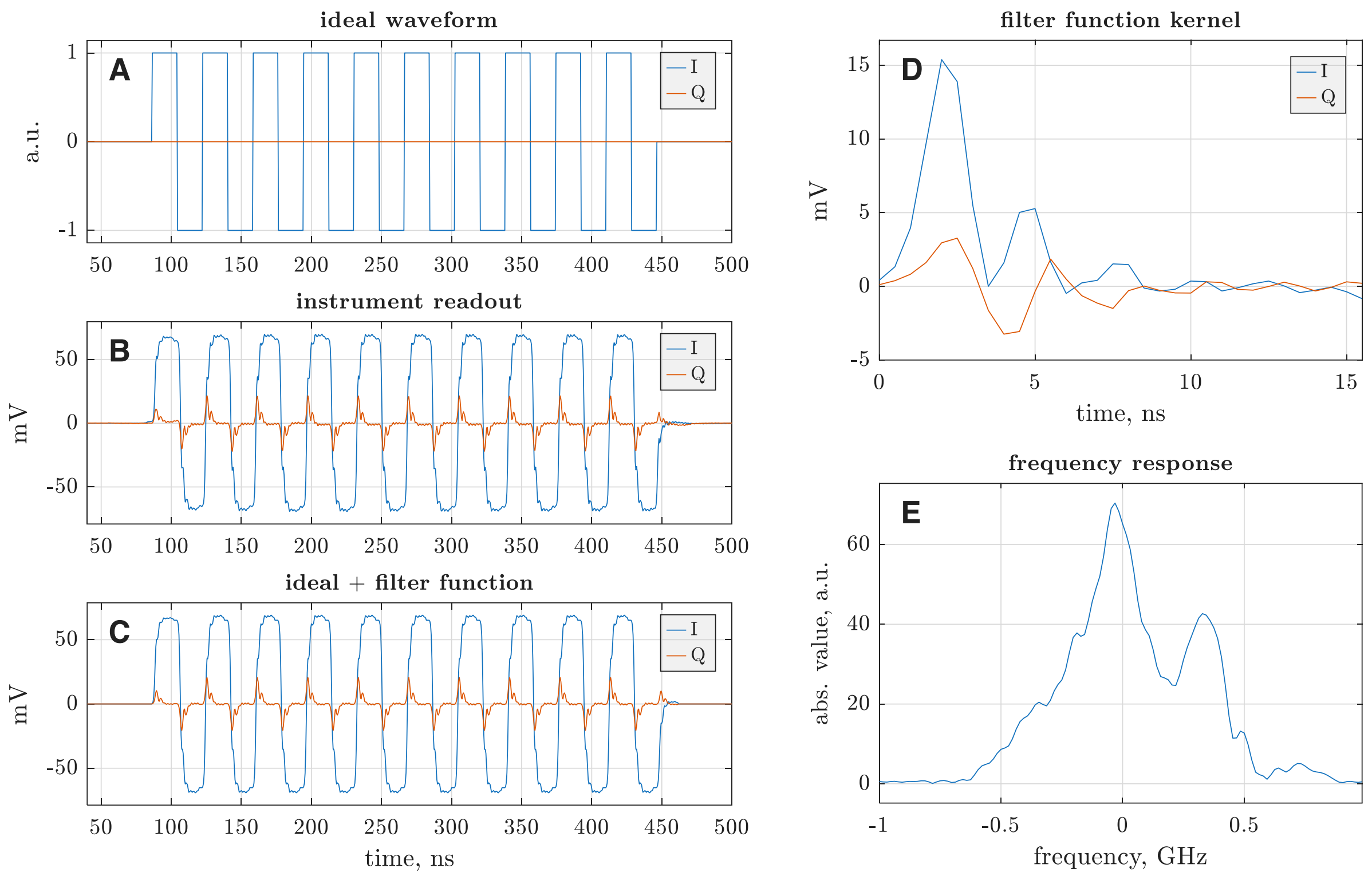

***Figure S4:*** *(A) Ideal X-inverse-X pulse sequence (the phases of the pulses alternate between 0° and 180° in 18 ns blocks) as programmed into the AWG. (B) HiPER's response at 94.000 GHz; real and imaginary components observed via the I and Q channels (Figure 2, Figure S1). (C) Convolution of the ideal XiX sequence with the extracted filter function. (D) Filter function kernel (in-phase and out-of-phase components), extracted from the instrument readout. (E) Frequency-domain representation of the filter function kernel.*

# S5. Amplitude shaping of linear frequency sweeps

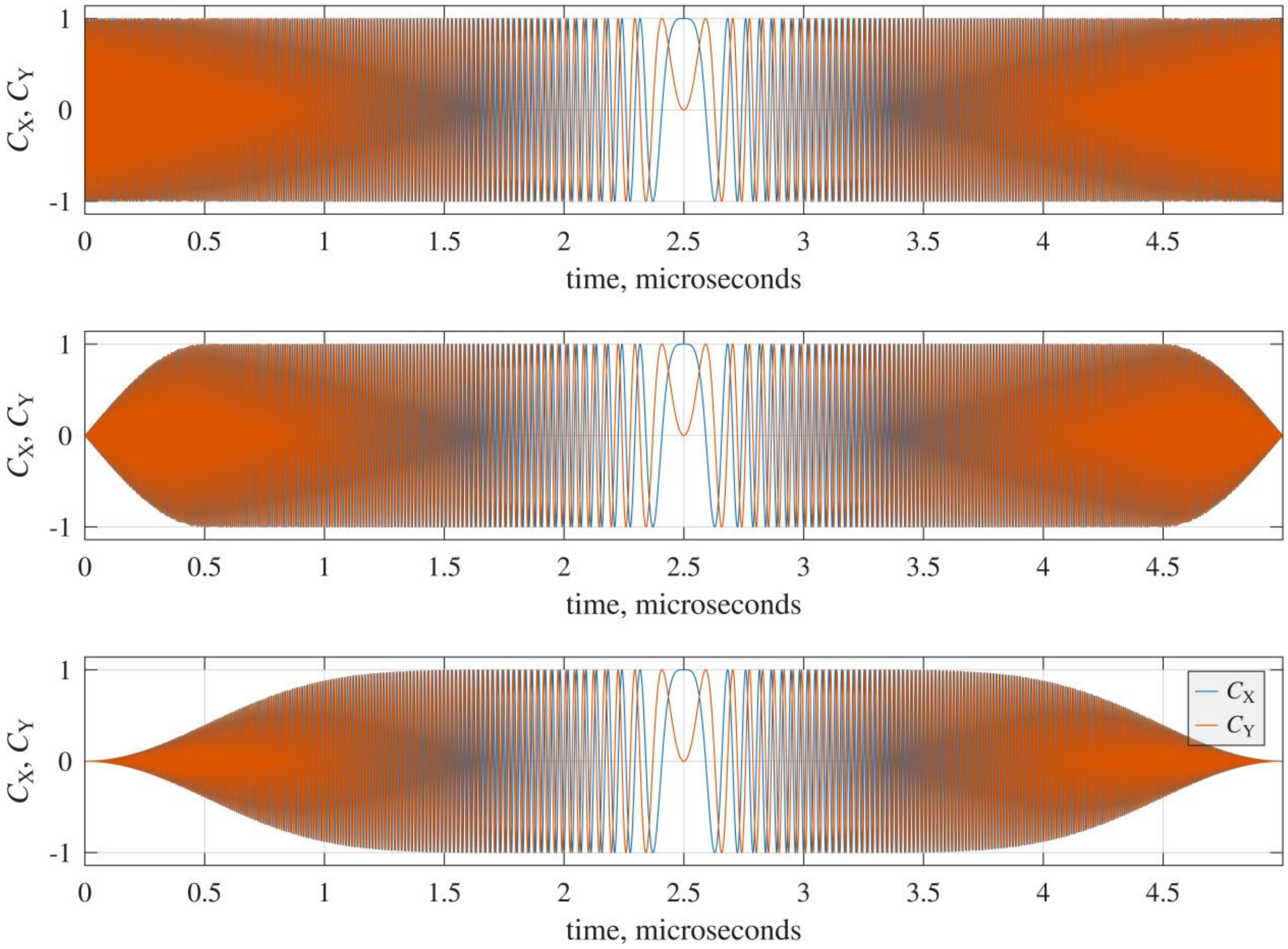

***Figure S5:*** *Real and imaginary components of a 300 MHz linear frequency sweep with constant amplitude (top), the amplitude tapered off at the edges using a quarter-sine wave (middle), and the amplitude tapered off at the edges following the wideband uniform rate smooth truncation (WURST)[4] profile (bottom).*

The amplitude of the quarter-sine wave tapered sweep at time $t$ is given by

$$a(t) = a_{\max} \times \begin{cases} \sin\left(\dfrac{n\pi t}{2T}\right) & \text{for} \quad 0 \le t \le \dfrac{T}{n} \\ 1 & \text{for} \quad \dfrac{T}{n} < t < \left(1 - \dfrac{1}{n}\right)T \\ \sin\left(\dfrac{n\pi}{2T}(T - t)\right) & \text{for} \quad \left(1 - \dfrac{1}{n}\right)T \le t \le T \end{cases}$$

where $T$ is the duration of the sweep (5 μs in this case) and $n$ is the smoothing factor (set here to 10). The amplitude profile of the WURST sweep (called sine bell power in Figure 5) is

$$a(t) = a_{\max}\left(1 - \left|\sin\left(\frac{\pi}{T}\left(t - \frac{T}{2}\right)\right)\right|^{n}\right)$$

with the smoothing $n$ factor also set to 10.

## S6. Contact curves of the integrated solid effect

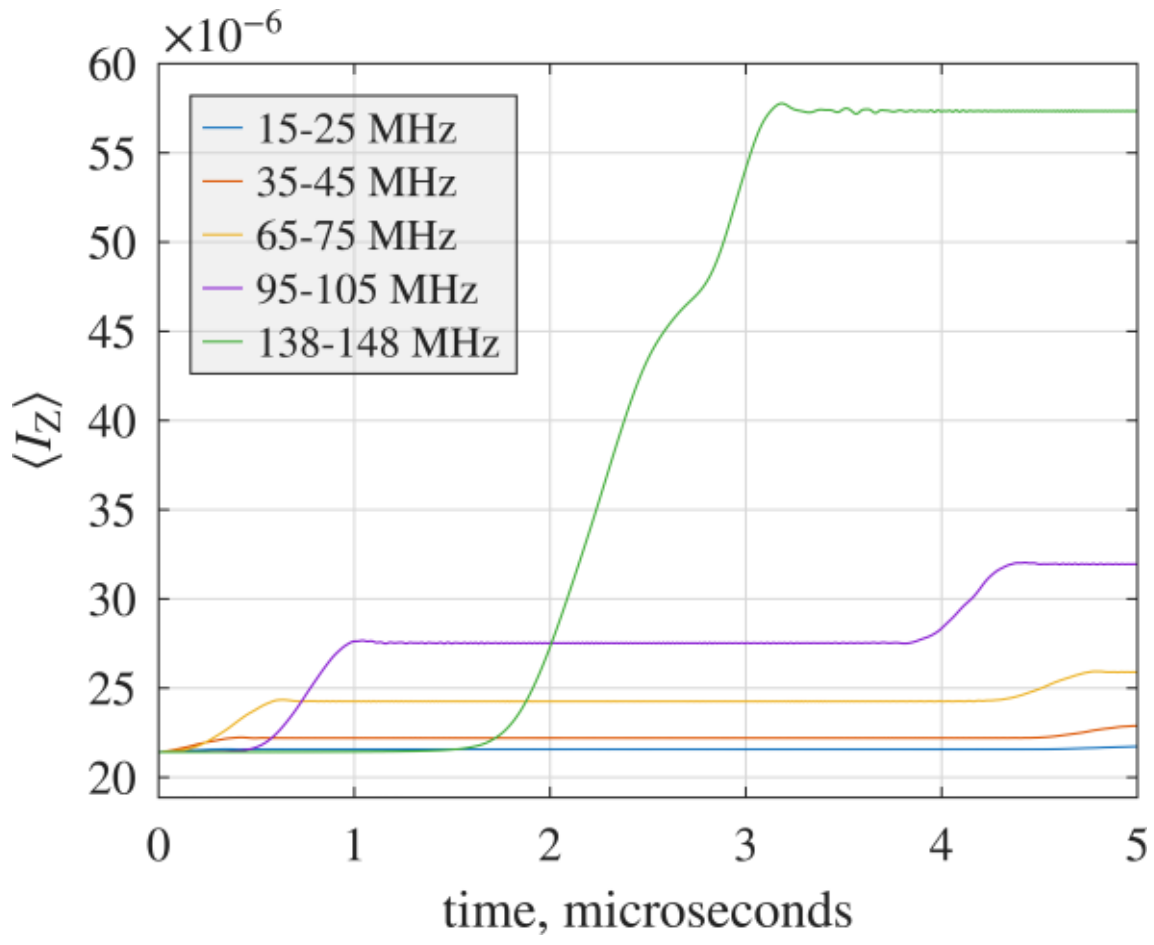

***Figure S6:*** *Single-shot transfer simulations showing the development of longitudinal proton magnetisation during a 300 MHz frequency sweep (duration 5 μs) centred at resonance (integrated solid effect condition). As the electron Rabi frequency increases, the adiabatic passages at the beginning and end of the sweep transfer magnetisation more efficiently. Eventually, when the electron Rabi frequency becomes approximately equal to the nuclear Larmor frequency (143 MHz), the two passages merge into one.*

## S7. Electron Rabi frequency distribution

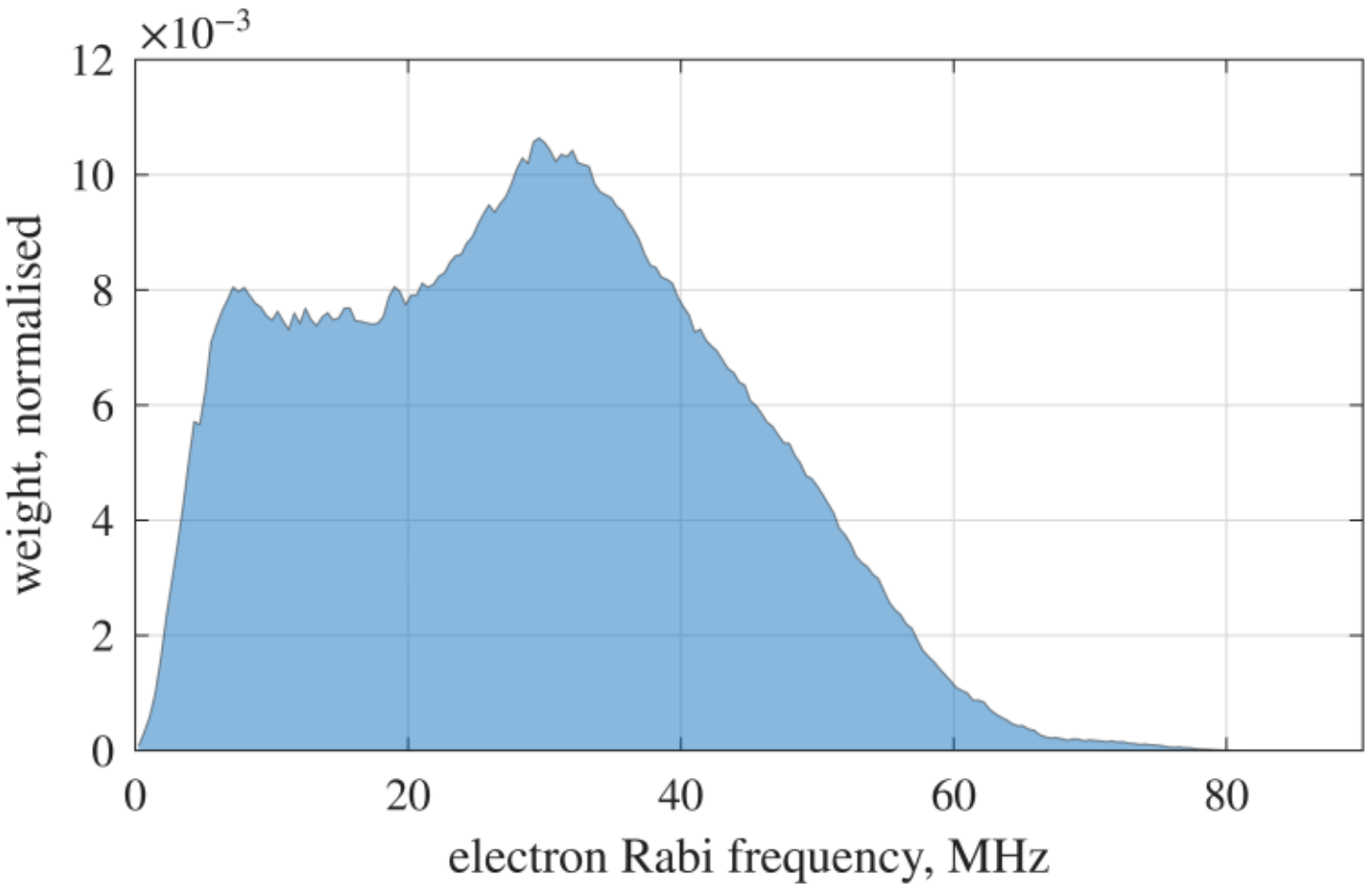

***Figure S7:*** *The standing-wave pattern in HiPER's sample holder has previously been analysed with finite element methods (CST Microwave Studio, Dassault Systèmes) by Zhao et al.*[5] *The predicted $B_1$ distribution, converted to electron Rabi frequencies, is plotted here. We estimate that, for an RVA setting of 3 dB, a peak microwave power of 650 W reaches the sample and assume an average effective conversion factor of 0.04 mT/√W. In simulations of pulsed DNP experiments performed with an RVA setting of 3 dB, this distribution is approximated by a uniform distribution between 8 and 42 MHz. In the design of DNP waveforms with GRAPE for experiments with an RVA setting of 7.5 dB (estimated peak power of 232 W), a uniform distribution between 5 and 25 MHz is assumed.*

## S8. Optimising integrated and stretched solid effect

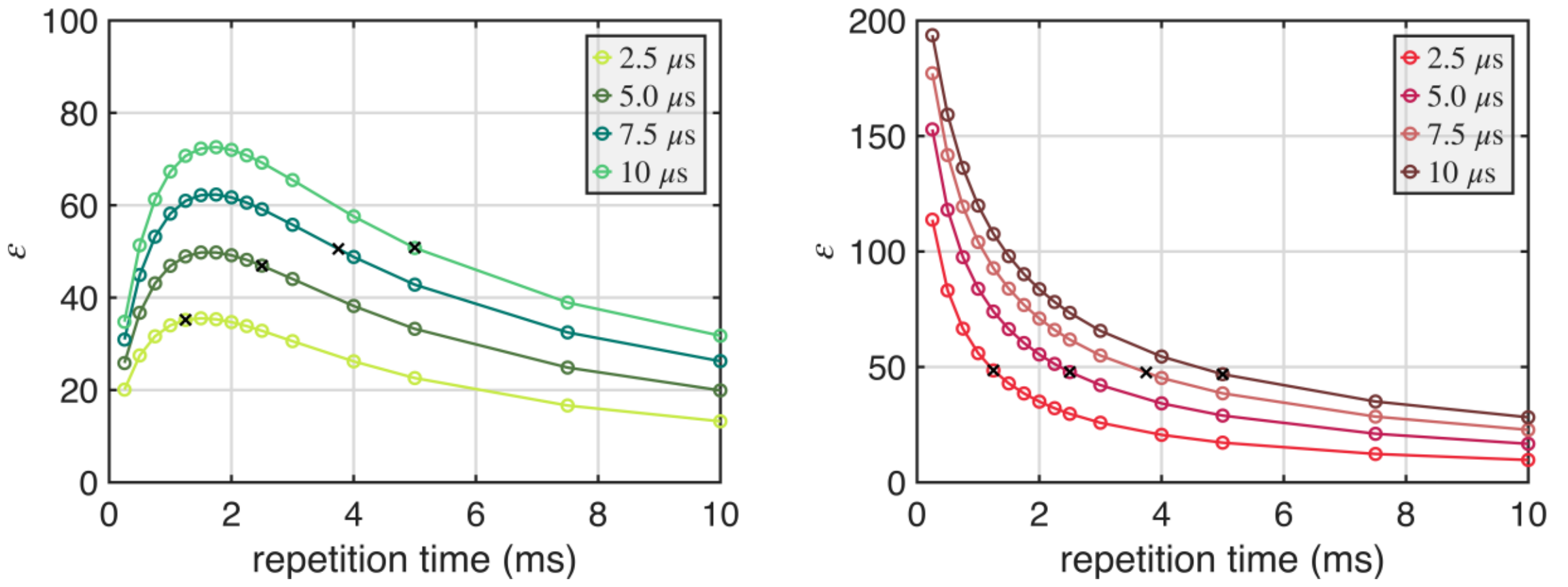

***Figure S8:*** *Steady state simulations illustrating the effects of pulse length and repetition time on the maximum enhancement obtained with 300 MHz frequency sweeps in pulsed DNP experiments at 3.35 T. On the left, enhancements are monitored on resonance (integrated solid effect). On the right, enhancements are monitored around a microwave resonance offset of 225 MHz (stretched solid effect). The distance ensemble (3.5-20 Å), the electron Rabi frequency ensemble (8-42 MHz, 3 dB RVA), and HiPER's filter function are accounted for in the simulation. The black crosses indicate an average applied microwave power of 1.3 W, as in the experiments of Figure 5 ($t_p$ = 5 μs) in the main text. The simulations show that, as the average power is increased, ISE goes through a maximum, whereas SSE continues to grow. This is in agreement with the predictions of Wenckebach and coworkers, based mostly on analytical derivations.[6,7] For ISE, lengthening the pulses increases the enhancement asymptotically.*

## S9. Rapid integrated solid effect: evolution and frequency profile

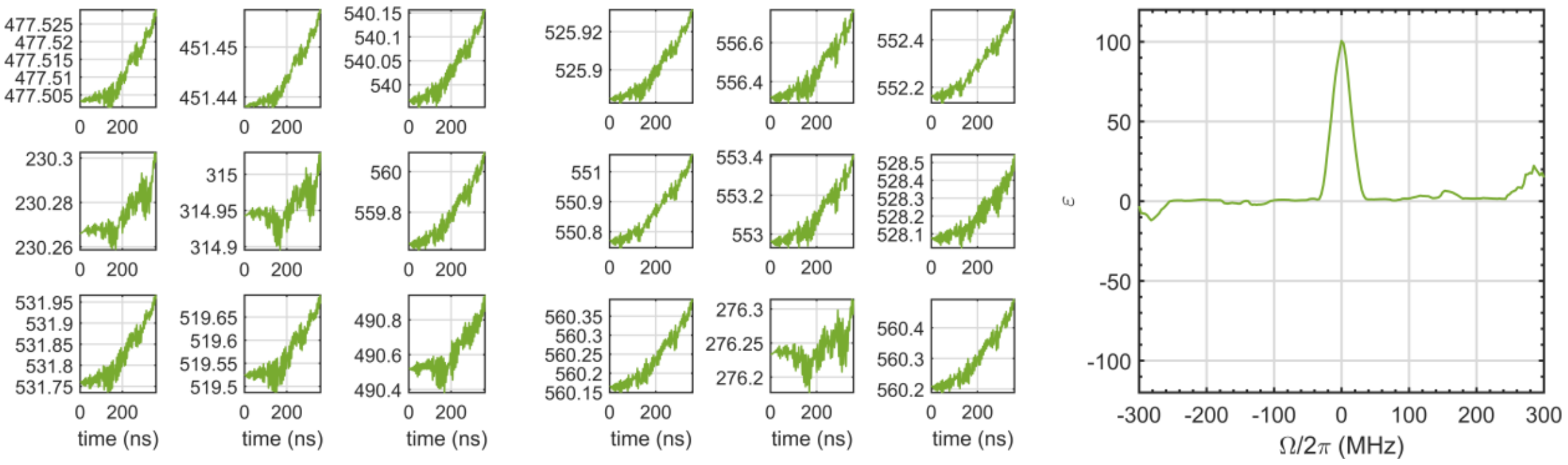

***Figure S9:*** *(left) Evolution of longitudinal proton magnetisation during the pulse part of the steady orbit of RISE for 18 randomly selected orientations of the 800-point REPULSION grid. (right) Steady-state simulation of the frequency profile of RISE, taking into account the distance ensemble (3.5-20 Å), the electron Rabi frequency ensemble of 5-25 MHz (associated with a 7.5 dB setting of the RVA), and HiPER's filter function.*

## S10. Adiabatic solid effect optimisation

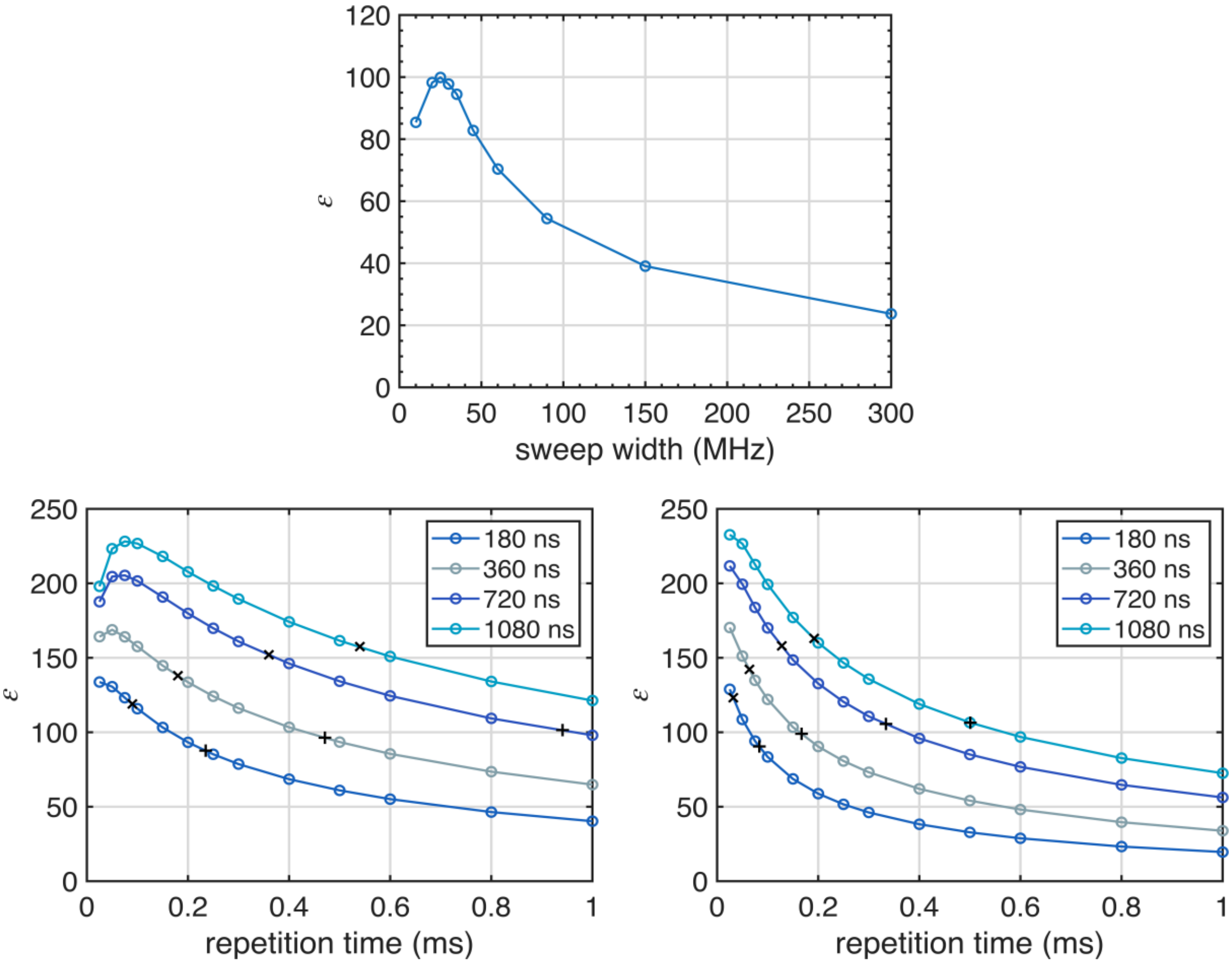

***Figure S10:*** *Steady-state simulations exploring the effects of sweep width, pulse duration, repetition time, and average power on the enhancements obtained with the adiabatic solid effect. All simulations included the distance ensemble (3.5-20 Å) and HiPER's filter function. (top) Optimisation of the sweep width. 25 MHz is optimal in the simulation; in experiments, 30 MHz gave the best results. The pulse duration was 360 ns, the repetition time was 167 μs, and a uniform electron Rabi frequency distribution of 5-25 MHz was included, matching experiments reported in Table 1 and Figures S12 and S13. Bottom row: optimisation of the repetition time for different pulse durations for electron Rabi frequency distributions of 8-42 MHz (bottom left) and 5-25 MHz (bottom right). The black crosses indicate an average power of 1.3 W (as in experiments of Figure 5), the black pluses indicate an average power of 0.5 W (as in experiments reported in Table 1). Compared to the ISE and the SSE, the narrow sweeps of the ASE allow more targeted polarisation transfer. Hence, higher enhancements are obtained, even at 2.8 times lower microwave irradiation power.*

# S11. HiPER-adapted DNP: evolution and frequency profiles

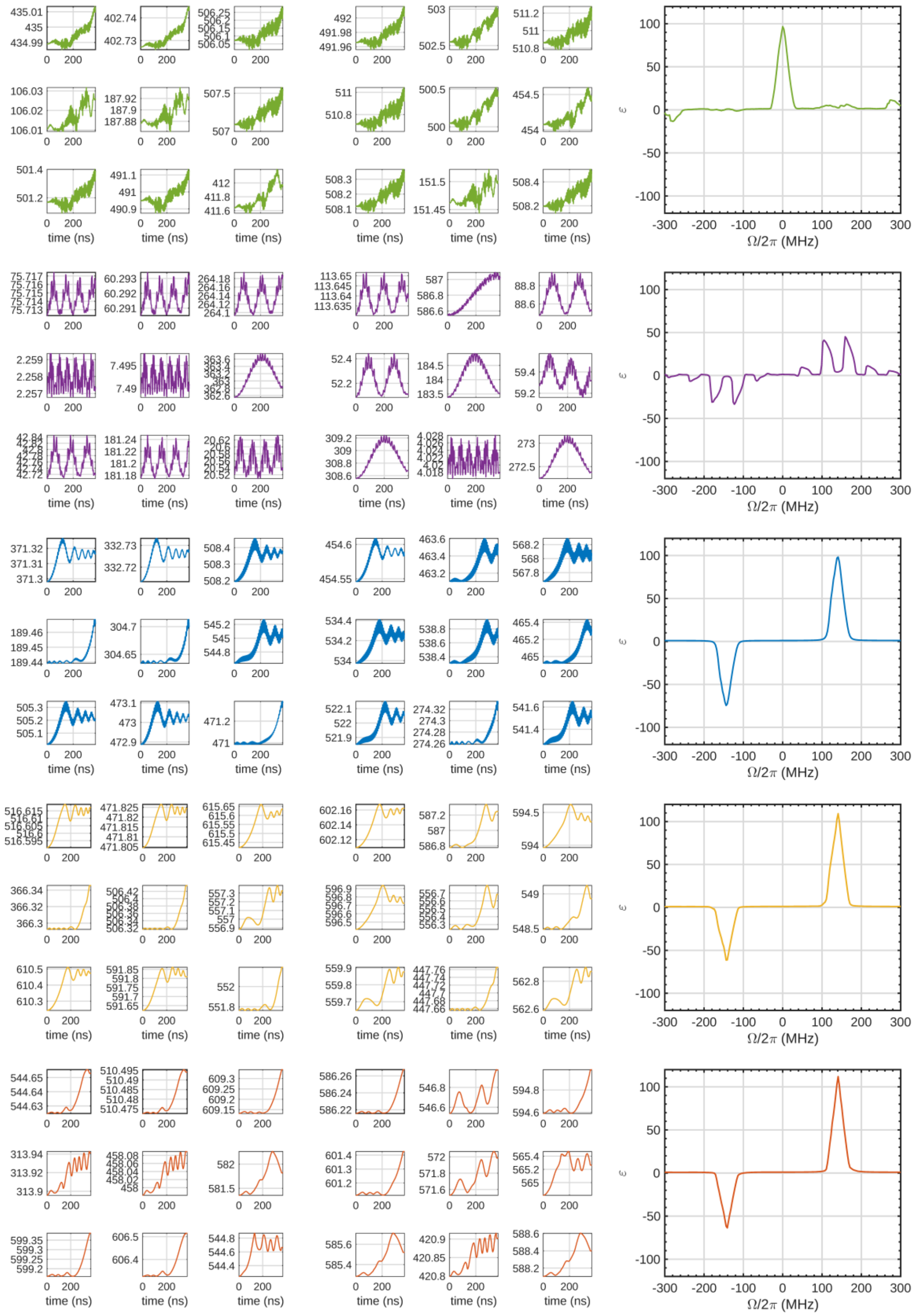

***Figure S11:*** *(left) Evolution of longitudinal $^1$H magnetisation during the pulse parts of the steady orbits of HiPER-adapted RISE, XiX, ASE, EASE, and ODQ for 18 randomly selected orientations of the 800-point REPULSION grid. (right) Steady-state simulations of frequency profiles of the same sequences. All simulations included the distance ensemble (3.5-20 Å), the Rabi frequency ensemble (5-25 MHz, 7.5 dB RVA), and the HiPER filter function.*

## S12. Offset optimisation and build-up curves

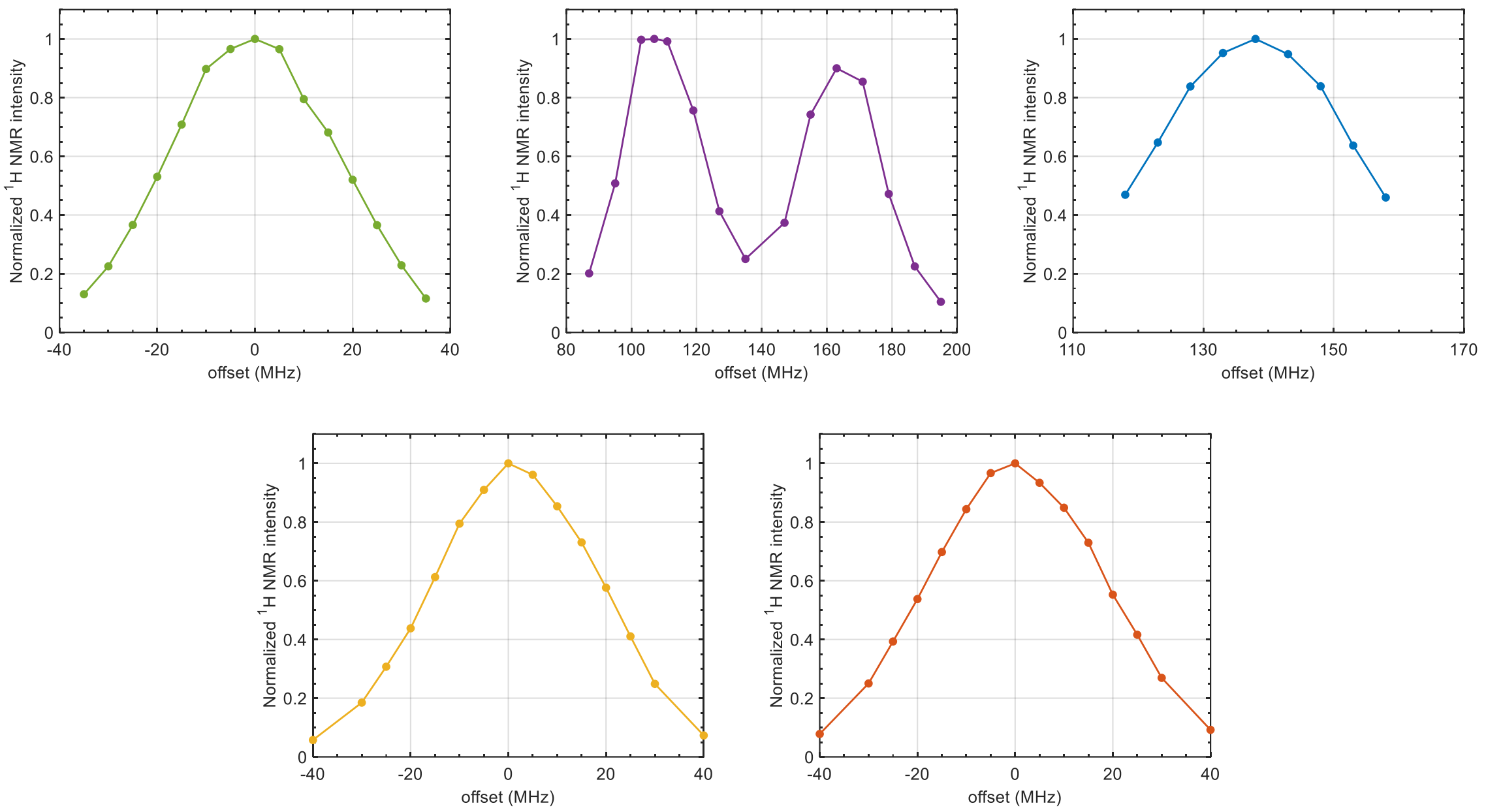

***Figure S12:*** *Experimental optimisation of the microwave resonance offsets (frequency profiles). Left to right and top to bottom: HiPER-adapted RISE (green), XiX (purple), ASE (blue), EASE (yellow), and ODQ (red). For ASE, the offset optimisation was performed with the RVA set to 7.5 dB. For the other DNP sequences, the RVA was set to 3 dB. Offsets are relative to 94.000 GHz. Note that HiPER-adapted RISE, EASE, and ODQ are optimized relative to 94.000 GHz, but frequency components are neither restricted to nor centred at this value, see the spectrograms in Figure 7 of the main manuscript.*

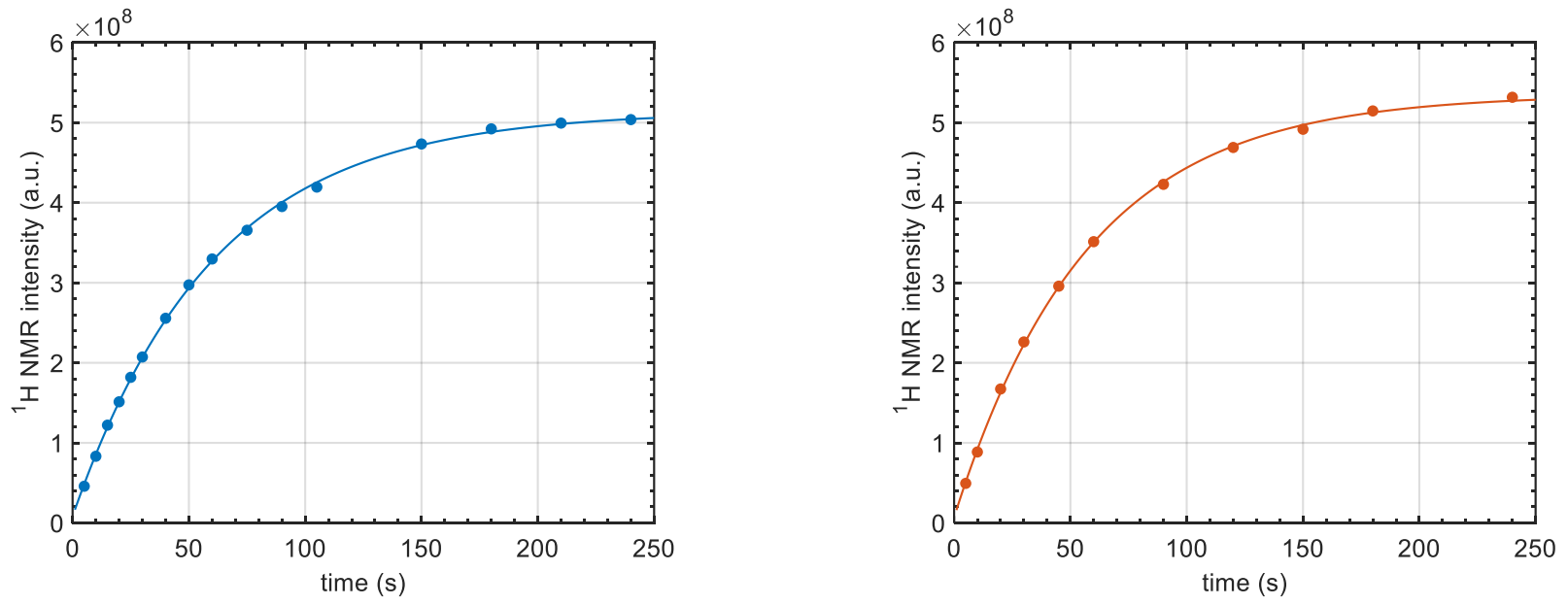

***Figure S13:*** *Experimental build-up of bulk proton magnetisation for (left) ASE and (right) ODQ. Characteristic build-up times are 60.2 s (standard error 0.9 s) and 57.1 s (standard error 1.1 s), respectively.*

# S13. Statistical analysis of DNP performance

***Table S1.*** *Measurements of $I_{on}$ (60 s of irradiation) used to characterise its repeatability. The scatter is proportional to the signal and its origin is most likely acquisition-to-acquisition variation in microwave power and sample temperature.*

| RVA att., dB | sequence | $I_{on}/10^8$ (repeats) | $T_{bu}$, seconds |
|---|---|---|---|
| 6.0 | RISE | 293, 280, 292 | - |
| | EASE | 381, 366, 378 | - |
| | ODQ | 397, 388, 386 | - |
| | ASE | 356, 379, 371 | - |
| 7.5 | ODQ | 336, 351 | 57.1 ± 1.1 |
| | ASE | 312, 330, 328 | 60.2 ± 0.9 |

The relevant quantity for assessing DNP performance is the practical sensitivity gain $G$ (Eq 15 in the main manuscript). With the enhancement factor written out, the expression is:

$$G = \frac{I_{\text{on}}}{I_{\text{off}}} \cdot \frac{\left(1 - e^{-d_{\text{off}}^{R}/T_{1\text{n}}}\right)}{\left(1 - e^{-d_{\text{on}}^{R}/T_{\text{bu}}}\right)} \cdot \sqrt{\frac{T_{1\text{n}}}{T_{\text{bu}}}} \quad (1)$$

The improvement provided by ODQ over ASE, I given by their practical sensitivity gain ratio:

$$\begin{aligned} R = \frac{G_{\text{ODQ}}}{G_{\text{ASE}}} &= \frac{I_{\text{on}}^{\text{ODQ}}}{I_{\text{on}}^{\text{ASE}}} \cdot \frac{1 - e^{-d_{\text{on}}^{\text{R}}/T_{\text{bu}}^{\text{ASE}}}}{1 - e^{-d_{\text{on}}^{\text{R}}/T_{\text{bu}}^{\text{ODQ}}}} \cdot \sqrt{\frac{T_{\text{bu}}^{\text{ASE}}}{T_{\text{bu}}^{\text{ODQ}}}} \\ &= \frac{343.5}{323.3} \times \frac{1 - e^{-60/60.2}}{1 - e^{-60/57.1}} \times \sqrt{\frac{60.2}{57.1}} = 1.058 \end{aligned} \quad (2)$$

Thus, ODQ is 5.8% stronger than ASE. A statistical treatment with propagation of uncertainties, detailed below, shows that this improvement is significant with the 95% confidence interval ranging from 0.4% to 11.5%.

Uncertainties were propagated through the ratio $R$. The quantities $I_{\text{off}}$ and $T_{1\text{n}}$ cancel, reducing the amplification of uncertainties. The irradiation time $d_{\text{on}}^{R} = 60$ s was treated as exact. Run-to-run variation in $I_{\text{on}}$ was described by a common relative standard deviation. Relative variances were pooled over the six repeat blocks in Table S1 (four at 6 dB and two at 7.5 dB):

$$s_{\text{rel}}^2 = \frac{\sum_g \left(n_g - 1\right)\left(s_g / \bar{I}_g\right)^2}{\sum_g \left(n_g - 1\right)} \quad (3)$$

Here $n_g$, $\bar{I}_g$, and $s_g$ are the number of repeats, mean intensity, and sample standard deviation in block $g$. The standard uncertainty of each mean was $u\left(\bar{I}_g\right) = s_{\text{rel}} \bar{I}_g / \sqrt{n_g}$. At 7.5 dB, ODQ had two repeats and ASE had three, giving relative standard uncertainties of 1.84% and 1.50%, respectively. The build-up-time uncertainties were taken directly from the fit standard errors: $T_{\mathrm{bu}}^{\mathrm{ODQ}} = 57.1 \pm 1.1$ s and $T_{\mathrm{bu}}^{\mathrm{ASE}} = 60.2 \pm 0.9$ s.

The four measured inputs were treated as independent. First-order propagation was performed in logarithmic space, using the sensitivities $S_x = \partial \ln R / \partial \ln x$:

$$u_{\ln R}^2 = \sum_x S_x^2 \left[ \frac{u(x)}{x} \right]^2, \qquad x \in \left\{ \bar{I}_{\mathrm{on}}^{\mathrm{ODQ}}, \bar{I}_{\mathrm{on}}^{\mathrm{ASE}}, T_{\mathrm{bu}}^{\mathrm{ODQ}}, T_{\mathrm{bu}}^{\mathrm{ASE}} \right\} \tag{4}$$

The intensity sensitivities are $+1$ and $-1$. The build-up time sensitivities are $+0.065$ (ODQ) and $-0.083$ (ASE): the steady state extrapolation and the square root contributions largely cancel. The resulting relative standard uncertainty is 2.38%, corresponding to $u(R) = 0.025$; the build-up-time terms account for less than 1% of the variance. A Monte Carlo calculation with $2 \times 10^6$ independent normal draws, using the same input means and standard uncertainties, confirmed the accuracy of the linearisation. Treating the intensity contributions as sharing one pooled variance estimate and neglecting the small build-up time terms in the degrees-of-freedom denominator then yielded $I_{\mathrm{ODQ}} / I_{\mathrm{ASE}} = 1.058$ with a 95% confidence interval from 0.4% to 11.5%. A two-sided test of $R = 1$ gave $p = 0.036$; the uncertainty is dominated by repeatability of $I_{\mathrm{on}}$.